\documentclass[conference]{IEEEtran}
\IEEEoverridecommandlockouts
\usepackage{cite}
\usepackage{amsmath,amssymb,amsfonts}
\usepackage{algorithmic}
\usepackage{graphicx}
\usepackage{textcomp}
\usepackage{xcolor}
\usepackage{subfig}
\usepackage[most]{tcolorbox}
\usepackage{url}
\usepackage{tagging}
\usepackage{booktabs}

\newtcolorbox{mybox}{
enhanced,
boxrule=0pt,frame hidden,
borderline west={4pt}{0pt}{green!75!black},
colback=green!10!white,
sharp corners
}

\def\BibTeX{{\rm B\kern-.05em{\sc i\kern-.025em b}\kern-.08em
    T\kern-.1667em\lower.7ex\hbox{E}\kern-.125emX}}
\begin{document}

\title{More for Less: Integrating Capability-Predominant and Capacity-Predominant Computing
\thanks{}
}

\author{\IEEEauthorblockN{Zhong Zheng}
\IEEEauthorblockA{\textit{Department of Computer Science} \\
\textit{University of Illinois Chicago}\\
zzheng33@uic.edu}
\and
\IEEEauthorblockN{Michael E. Papka}
\IEEEauthorblockA{\textit{Argonne National Laboratory} \\
\textit{University of Illinois Chicago}\\
papka@uic.edu}
\and
\IEEEauthorblockN{Zhiling Lan}
\IEEEauthorblockA{\textit{University of Illinois Chicago} \\
\textit{Argonne National Laboratory}\\
zlan@uic.edu}
}

\maketitle
\begin{abstract}
Capability jobs (e.g., large, long-running tasks) and capacity jobs (e.g., small, short-running tasks) are two common types of workloads in high-performance computing (HPC). Different HPC systems are typically deployed to handle distinct computing workloads. For example, Theta at the Argonne Leadership Computing Facility (ALCF) primarily serves capability jobs, while Cori at the National Energy Research Scientific Computing Center (NERSC) predominantly handles capacity workloads. However, this segregation often leads to inefficient resource utilization and higher costs due to the need for operating separate computing platforms. This work examines what-if scenarios for integrating siloed platforms. Specifically, we collect and characterize two real workloads from production systems at DOE laboratories, representing capability-predominant and capacity-predominant computing, respectively. We investigate two approaches to unification. Workload fusion explores how efficiently resources are utilized when a unified system accommodates diverse workloads, whereas workload injection identifies opportunities to enhance resource utilization on capability computing systems by leveraging capacity jobs. Finally, through extensive trace-based, event-driven simulations, we explore the potential benefits of co-scheduling both types of jobs on a unified system to enhance resource utilization and reduce costs, offering new insights for future research in unified computing.
\end{abstract}


\section{Introduction}\label{Introduction}

\textbf{Motivation.} Historically, \emph{capability computing} and \emph{capacity computing} are two different paradigms in HPC, each serving distinct purposes. Capability computing, focused on providing the maximum processing power to solve complex, large-scale simulations, has been designed to handle highly demanding computational applications that require massive processing capabilities. On the other hand, capacity computing focuses on providing scalable and cost-effective computing resources supporting many concurrent applications. It is typically geared towards efficiently managing workloads with varying resource requirements. This dichotomy has shaped the HPC landscape, catering to the specific needs of diverse applications. 

However, the historical separation between capability and capacity computing has become less clear as the systems begin to look more and more like each other. The differentiation between sophisticated capability and cost-effective capacity computing resources diminishes as both converge toward employing comparable hardware resources. For example, similar processing devices and networking configurations (e.g., Cray XC40 using Dragonfly topology) were utilized to deploy capability computing and capacity computing platforms at DOE labs \cite{theta,cori}. 
As the hardware and software stack converge for capability and capacity computing, the primary distinction lies in the various scheduling policies employed to manage different user jobs. Capability computing prioritizes large and long-running jobs, whereas capacity computing frequently adopts a first-come, first-serve scheduling policy. The distinct nature of these paradigms has led to silos in the computing ecosystem, creating barriers to higher resource utilization and wider accessibility of HPC resources for broader applications. For example, a capability computing system could suffer from resource underutilization due to the lack of workloads with various resource requirements or job duration. Similarly, a capacity computing system could become overloaded due to the temporal bursty nature of user jobs.

Recognizing the limitations of the separate paradigms, several studies have proposed various ways to address the silo issues. These include designing different scheduling policies, migrating high-throughput jobs to capability systems, or utilizing cloud bursting to handle increasing workloads 
\cite{bicer2011framework,nair2010towards,fan2019scheduling,mao2016resource,rlscheduler,gunasekaran2019spock,shahrad2020serverless,farahabady2013pareto}. 
Despite these efforts, there are still many unexplored questions. 
In particular, open problems, such as resource utilization and cost-efficiency of separate computing infrastructures, highlight \emph{the need for a more integrated approach.}

\textbf{Objective.} In this work, we present a quantitative analysis of unifying capability-predominant and capacity-predominant computing. Capability-predominant computing refers to systems that have a minimum requirement for job size; for example, on the Theta system at ALCF, the minimum node requirement is 128. In contrast, capacity-predominant computing is characterized by systems that primarily handle small-sized jobs, such as Cori at NERSC, where over 96\% of jobs use no more than 32 computing nodes (more details in Section~\ref{charact}).

The goal of this analysis is to explore the potential benefits of creating a unified and versatile computing infrastructure. This unification could bridge the gap between capability simulations and capacity computing, providing a cost-effective environment for handling diverse workloads. By consolidating resources onto a unified platform, we can reduce redundant operational expenses, such as system maintenance and software licensing, and facilitate better resource utilization, thereby achieving the goal of \emph{more for less}

\textbf{Challenges.} Conducting such an integration analysis poses several key challenges.
First, while real-world experiments offer the most accurate understanding of system performance, building a comparable unified system and requiring all users to re-execute their jobs on it is not feasible. Hence, trace-based modeling is indispensable. 
Unfortunately, access to production workload traces is often limited, primarily due to the sensitive nature of the data. Supercomputers at HPC facilities often handle classified or proprietary research, making organizations cautious about sharing workload traces to prevent the disclosure of sensitive information.
Finally, analyzing workload traces, even when accessible, presents additional challenges due to the large volumes of data, necessitating advanced simulation techniques. Thus, integrating workload traces with high-fidelity simulations is essential for gaining insights and exploring the potential of a unified infrastructure.

\textbf{Key Contributions.} This work aims to tackle the above challenges. Specifically, our work highlights three key aspects.
First, we collect two representative workloads from widely used production systems at DOE HPC facilities during the same timeframe (2022). One is from the Theta machine at Argonne Leadership Computing Facility (ALCF), representing capability-predominant computing \cite{theta}. The other is from the Cori machine at the National Energy Research Scientific Computing (NERSC), representing capacity-predominant computing\cite{cori}. We perform \emph{a detailed workload characterization} and emphasize the key features of both workloads.

Second, we enhance an open-source scheduling simulator and provide an in-depth predictive analysis of integrating two types of computing paradigms through trace-based, event-driven scheduling simulation. The enhanced simulator lets us explore what-if scenarios, validate hypotheses, and predict system behavior. 
Specifically, we investigate two unification scenarios. The first is \emph{workload fusion}, where we explore a unified system for handling the consolidated Cori and Theta jobs to assess its impact on resource utilization and user experience as the system size decreases. The second scenario, \emph{workload injection}, entails transferring selected capacity jobs from Cori to Theta to utilize idle or wasted resources on the capability system. We analyze the effect of injecting additional capacity computing jobs on the capability computing jobs.

Third, our integration analysis highlights the potential benefits and impacts of consolidating different types of jobs on a unified system. By analyzing how diverse workloads behave on a unified system with varying reduced capacities, we identify opportunities to improve resource utilization while maintaining a satisfactory user experience and achieving cost savings.
The insights gained from the workload fusion and workload injection analyses \emph{provide a foundation for future research in unified computing}, such as planning resource allocations or expansions and aligning the computing infrastructure with the evolving needs of the user community.



\textbf{Paper Outline.}  The rest of the paper is organized as follows. Section \ref{Background} introduces the related work and background. Section \ref{charact} characterizes capability-predominant and capacity-predominant workloads. Section \ref{method} describes the integration methods. Section \ref{fusion} and Section \ref{injection} present the workload fusion and workload injection analyses, followed by the conclusion in Section \ref{Conclusion}. 

\section{Background and Related Work}\label{Background}
\subsection{Capability Computing and Capacity Computing}
HPC facilities are often specialized for either \emph{capability computing} or \emph{capacity computing}. Facilities like the National Energy Research Scientific Computing Center (NERSC) are dedicated to capacity computing, providing cost-efficient systems to effectively manage diverse workloads with varying resource requirements, including moderately large and numerous small problems \cite{national2008potential}. 
First-Come First-Serve (FCFS) with EASY backfilling is commonly used in scheduling at capacity computing facilities. FCFS sorts jobs in the wait queue by their arrival times and executes jobs from the head of the queue. When resources are insufficient for the first job, backfilling is triggered. The scheduler reserves the estimated starting time and required resources for the head job and then selects the jobs in the queue whose executions do not delay the head job, known as EASY backfilling\cite{mu2001utilization}.

On the other hand, facilities like the Argonne Leadership Computing Facility (ALCF) are specifically geared towards capability computing, emphasizing the utilization of maximum computing power to tackle large-scale problems efficiently in the shortest possible time\cite{national2008potential,allcock2017experience}. 
At ALCF, a utility-based scheduling policy named WFP, is deployed to support the mission of running large-scale capability jobs. WFP periodically calculates a priority score $\left( \frac{t_{queue}^2}{t_{supplied}^3} \right) \times \frac {n}{N}$ for each waiting job, where $t_{\text{queue}}$, $t_{\text{supplied}}$, $n$, and $N$ denote job wait time, user-supplied runtime,  job size, and machine size, respectively. The facility can adjust the score in certain cases \cite{allcock2017experience}. EASY back-filling is also deployed for boosting resource utilization\cite{mu2001utilization}.

\begin{figure}
    \centering
    \includegraphics[width=0.8\linewidth]{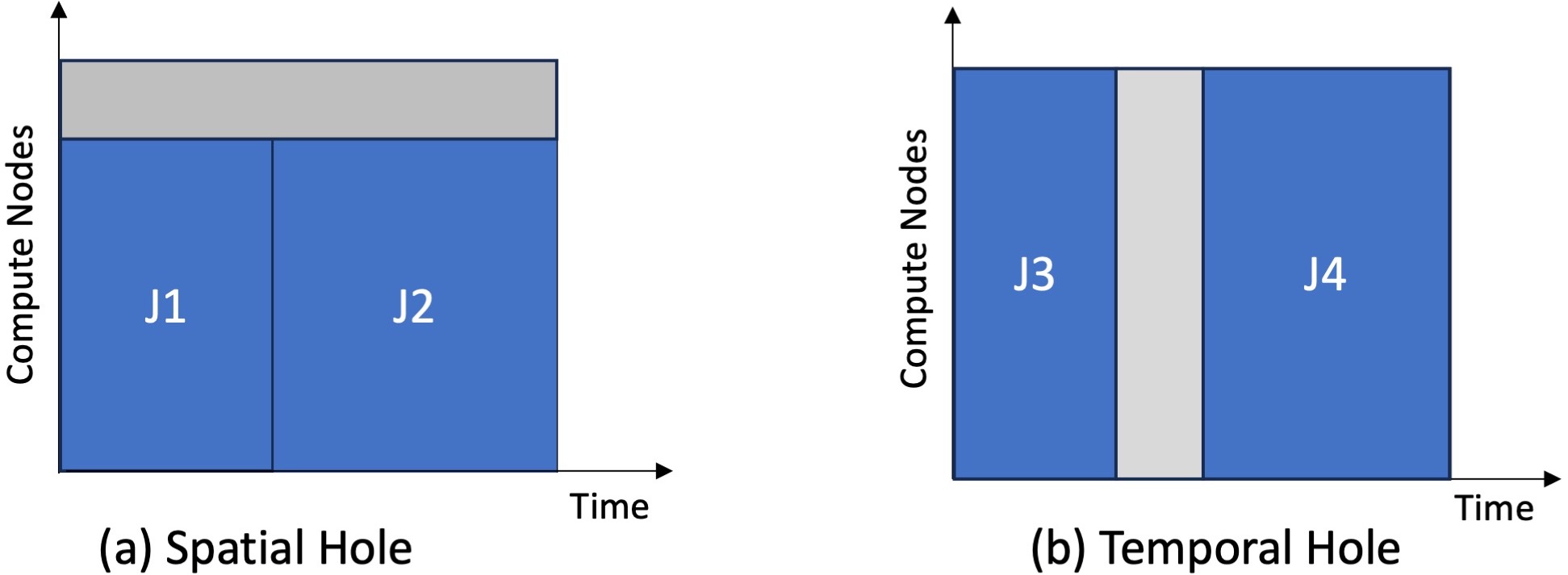}
    \caption{\small{Illustrative resource waste (gray area) in a capability system: (a) spatial hole due to the lack of small-sized jobs; (b) temporal hole due to the lack of short-running jobs.}}
    \label{fig:scheduling hole}
    \vspace{-20pt}
\end{figure}

Although the specialization allows each facility to tailor its infrastructure to excel in either capability or capacity computing domains, the separation between capability and capacity computing has not been without its problems. One challenge is resource underutilization. We have observed two types of resource wastage cases in our study. Figure \ref{fig:scheduling hole} provides illustrative examples to demonstrate them: (a) \emph{spatial hole} and (b) \emph{temporal hole}. In (a), J1 and J2 occupy the majority of the computing nodes for a given period, leaving only a small chunk of computing resources. Capability jobs are too large to be executed in a small chunk, leading to a spatial hole. In (b), there is a short period between the execution of J3 and J4, and this temporal hole cannot be filled by any long-running job in the wait queue. 

\subsection{Related Studies}
\label{sec:2}
For capacity and capability computing, resource utilization is a crucial metric that measures the extent to which resources are effectively employed in performing tasks \cite{emeras2014analysis}. Researchers point out that large-scale clusters are experiencing resource underutilization 
\cite{you2013comprehensive,rodrigo2018towards,cortez2017resource,li2023analyzing,patel2020job,ferreira2021workflows,li2022ai}.   
Recognizing the constraints inherent in separate paradigms, various studies have presented different strategies to overcome silo issues. These strategies involve (1) the enhancement of scheduling policies, (2) the migration of high-throughput jobs or Function-as-a-Service (FaaS) jobs to capability computing systems, and (3) the adoption of cloud bursting to address escalating workloads. 

To increase resource utilization at the system level, efforts have primarily focused on enhancing scheduling policies. Some studies adopt optimization algorithms to approximate optimal scheduling decisions concerning achieving higher utilization. For instance, Zheng et al. \cite{plan_based} present a plan-based scheduling algorithm, adopting simulated annealing as the optimization engine to improve the system utilization. Fan et al. \cite{fan2019scheduling} explore the multi-objective optimization to maximize computing node utilization with burst buffer utilization. Recent efforts begin to leverage reinforcement learning (RL) for resource management and job scheduling \cite{mao2016resource,rlscheduler, fan2021deep}.

While improving scheduling policies can boost resource utilization in large-scale clusters, node idleness often persists, as these idle slots are typically too short or small for any HPC job. Hence, utilizing small and short computing tasks, such as high-throughput or function-as-a-service jobs, becomes a viable option for addressing this issue. Various studies have explored the feasibility of migrating high-throughput computing (HTC) jobs or FaaS into HPC computing systems to minimize resource fragmentation \cite{gunasekaran2019spock,przybylski2022using,shahrad2020serverless,roy2022mashup,du2019feasibility,lahiff2020running}. For example, Du et al. \cite{du2019feasibility} propose a mechanism using HTCondor to facilitate the migration of high-throughput jobs between HTCondor and Slurm systems. Przybylski et al. \cite{przybylski2022using} develop a FaaS infrastructure and deploy the FaaS invokers on idle HPC cluster nodes, allowing FaaS to utilize these idle resources with minimal impact on HPC job performance. 

Contrasting with the resource underutilization in capability computing systems, capacity computing systems often experience excessively high demands of resource requests \cite{du2019feasibility}. Various studies have addressed this issue by adopting cloud bursting to harness public cloud computing whenever the on-premises infrastructure approaches its peak capacity\cite{bicer2011framework,farahabady2013pareto,nair2010towards,guo2014cost, guo2012seagull}. For instance, PANDA is a framework that incorporates a Fully Polynomial-Time Approximation Scheme (FPTAS) to schedule computing tasks across resources allocated in both private and public clouds, addressing excessive resource requirements \cite{farahabady2013pareto}.

\emph{This study distinguishes from these studies in several aspects}. First, it does not intend to develop new scheduling policies to boost either capability or capacity computing. Instead, it investigates improving resource utilization by co-running capability and capacity jobs within a unified system. Complementary to the studies of leveraging HTC or FaaS jobs for capability computing, this work emphasizes accommodating capacity jobs on a capability platform. While integrating HTC or FaaS jobs typically requires additional runtime support, our integration approach simplifies the process by solely relying on the selection of a scheduling policy without the need for extra software assistance. Finally, unlike the cloud bursting studies, our work explores using resources already owned as a unified system.

\section{Workload Characterization} 
\label{charact}

\begin{table}
\caption{Theta and Cori Workloads}
\vspace{-5pt}
\label{Theta and Cori Workloads}
\small
    \begin{tabular}{|c|p{2.2cm}|p{2.2cm}|} \hline  
         &  \textbf{Theta}& \textbf{Cori}\\ \hline  
         Location&  ALCF& NERSC\\ \hline  
         Category&  Capability-predominant & Capacity-predominant \\ \hline  
         Scheduler& Cobalt & Slurm \\ \hline
         Machine Type& Cray XC40 & Cray XC40 \\ \hline
         Compute Nodes &  4,360 & 9,688 \\ \hline  
         Processor & Intel Xeon Phi 7230 64-core & Intel Xeon Phi 7250 68-core \\ \hline
         Interconnect & Dragonfly & Dragonfly \\ \hline
         Trace Period&  1/1/2022- 12/31/2022& 1/1/2022-12/31/2022\\ \hline  
         Number of Jobs&  23,911& 2,349,370\\ \hline  
         Min Job Size&  128 nodes& 1 node\\ \hline 
    \end{tabular}
    \vspace{-15pt}
\end{table}

We first characterize representative capacity-predominant and capability-predominant computing with several important takeaways, which form guiding principles for the integration analyses presented in Sections~\ref{method}-~\ref{injection}.

We collected and analyzed two production workloads: one from Theta at ALCF \cite{theta} and the other from Cori at NERSC \cite{cori}, both spanning a year during the same timeframe. Table~\ref{Theta and Cori Workloads} provides details of these logs. These machines were selected because they have similar hardware and networking technologies, yet they serve different workloads during the same period.

Theta consists of 4,392 Intel Xeon Phi Knights Landing (KNL) processors \cite{KNL}. For a fair analysis, we removed 32 debugging nodes and corresponding debugging jobs. Cobalt was used for batch scheduling on Theta, prioritizing large or long-waiting jobs using a utility function known as WFP (Section \ref{sec:2}), along with EACY backfilling.  On Theta, the minimum job size is 128 nodes.

Cori has 2,388 Intel Haswell processors \cite{Haswell} and 9,688  Intel Xeon Phi Knights Landing processors. To ensure fair experiments, we selected only Cori's KNL workloads for this study to ensure fair experiments. 
Slurm was used for batch scheduling on Cori, employing the FCFS policy with EASY backfilling. 

Figure~\ref{fig:job_distribution} provides a side-by-side comparison of the workloads in terms of job sizes and job runtimes. 
The plots clearly show that the Cori workload is dominated with small jobs, with 96\% of jobs using fewer than 32 nodes. Both workloads contain jobs with varying runtimes. An interesting observation is that over 10\% of the jobs in the Cori log have runtimes exceeding 60,000 seconds (16.7 hours).

\begin{mybox}
\textbf{Observation 1.} Both capability-predominant and capacity-predominant platforms can achieve high utilization, although utilization fluctuates. 
Interestingly, high usage on one system may coincide with low usage on another, suggesting potential workload balancing across platforms.
\end{mybox}

\begin{mybox}
\textbf{Observation 2.} Jobs on Cori tend to request significantly fewer computing resources and have shorter execution times compared to those on Theta. Over 75\% of Cori jobs request only one compute node, and 60\% of Cori jobs run for less than 1 hour.
\end{mybox}

\begin{mybox}
\textbf{Observation 3.} Jobs on Theta are large (more than 128 nodes) and long-running (e.g., 79\% of jobs run for over one hour). These capability jobs are more likely to create spatial and temporal gaps in scheduling, leading to relatively lower utilization and greater variance on Theta compared to Cori.
\end{mybox}

\begin{figure} 
    \centering
    \subfloat[Theta]{%
       \includegraphics[width=0.86\linewidth]{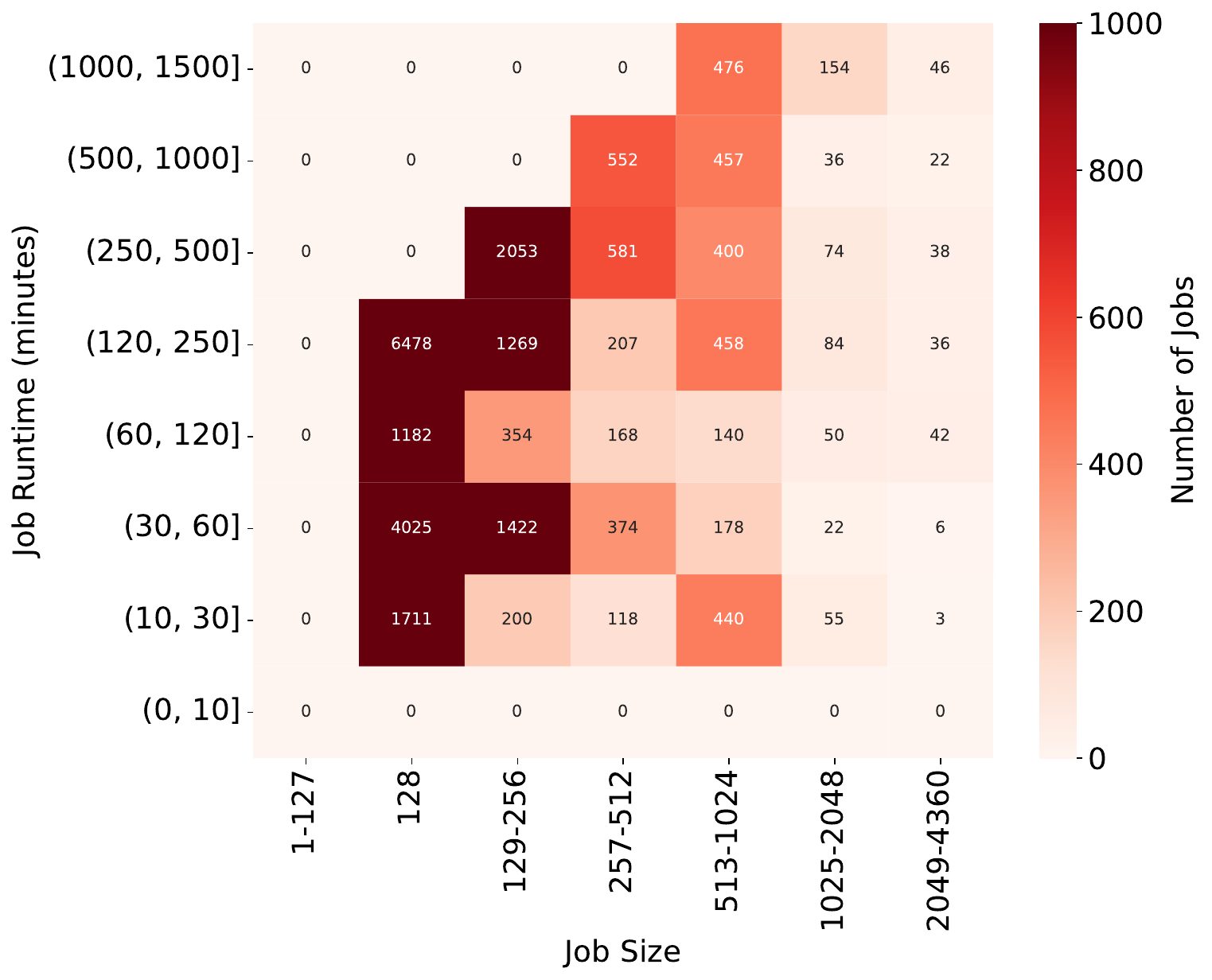}
       \label{fig:theta_dist}}\hfill
    \subfloat[Cori]{%
        \includegraphics[width=0.9\linewidth]{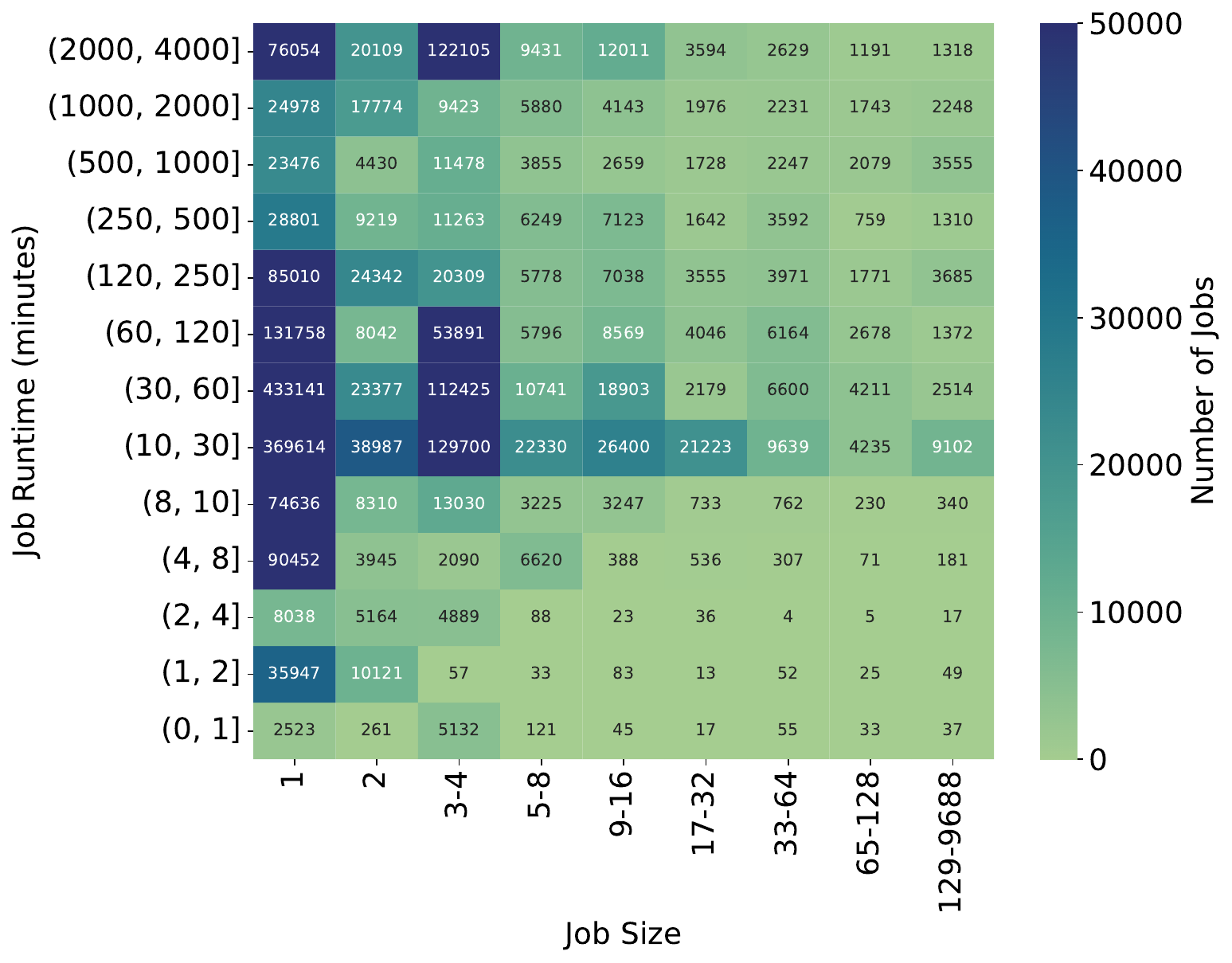}
        \label{fig:cori_dist}}
    \caption{Theta and Cori job distributions. }
    \vspace{-10pt}
    \label{fig:theta_cori_distribution}
\end{figure}

Figure \ref{fig:theta_cori_distribution} shows job distribution, characterized by both job size and runtime. 
Since Theta is a capability-predominant computing system typically running large-sized capability jobs, the minimum job size on Theta is set at 128 nodes. On the contrary, Cori accommodates a large number of single-node jobs. Interestingly, over 75\% of Cori jobs request only one computing node, and 73\% of these jobs require no more than 60 minutes for execution. Additionally, no Theta job has a runtime of less than 10 minutes, whereas numerous Cori jobs have a job runtime shorter than 10 minutes. 


\begin{figure} 
    \centering
    \subfloat[Job distribution by job sizes]{%
       \includegraphics[width=0.8\linewidth]{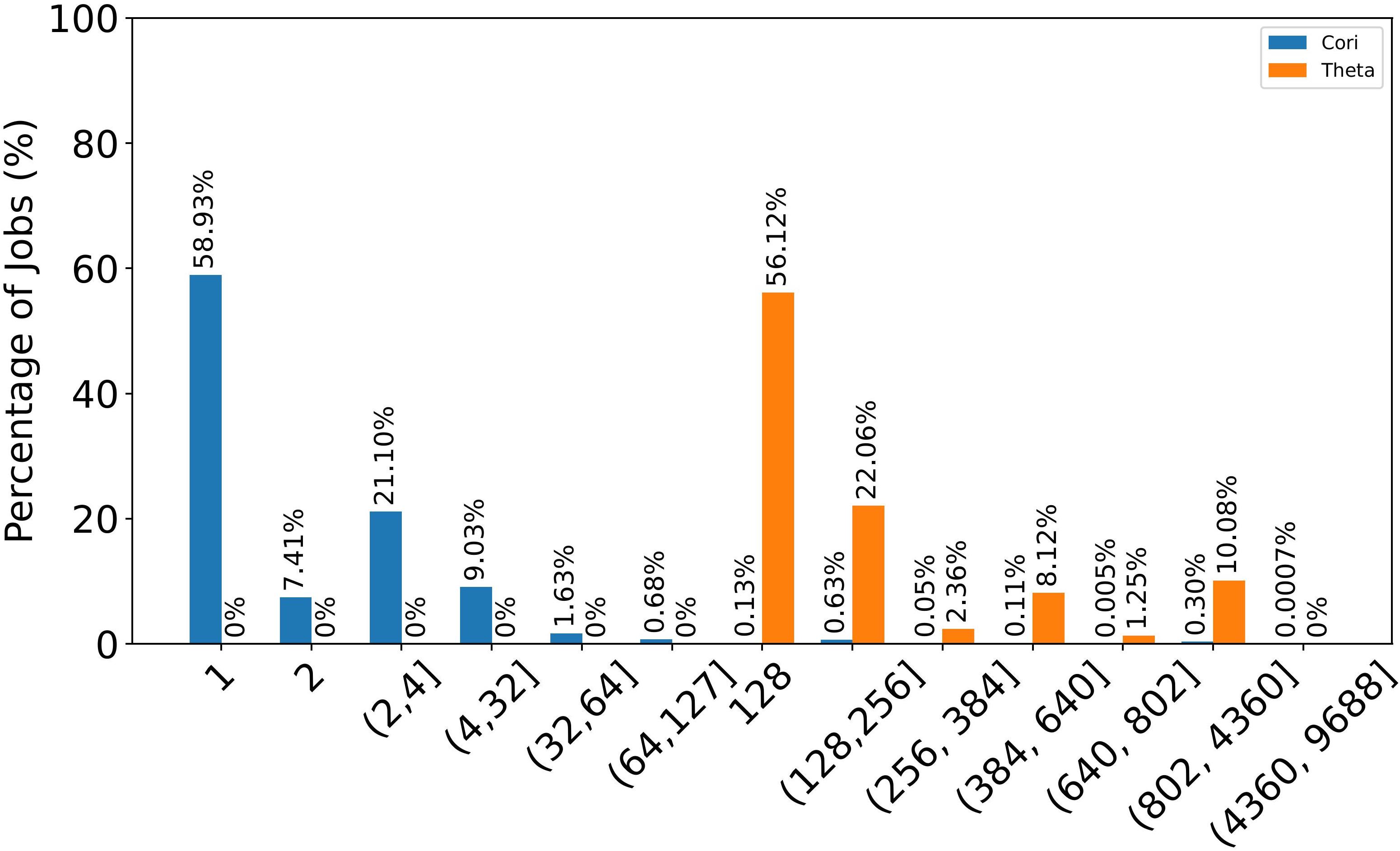}
       \label{fig:job_distribution}}\hfill
    \subfloat[Job distribution by job runtimes (in seconds)]{%
        \includegraphics[width=0.8\linewidth]{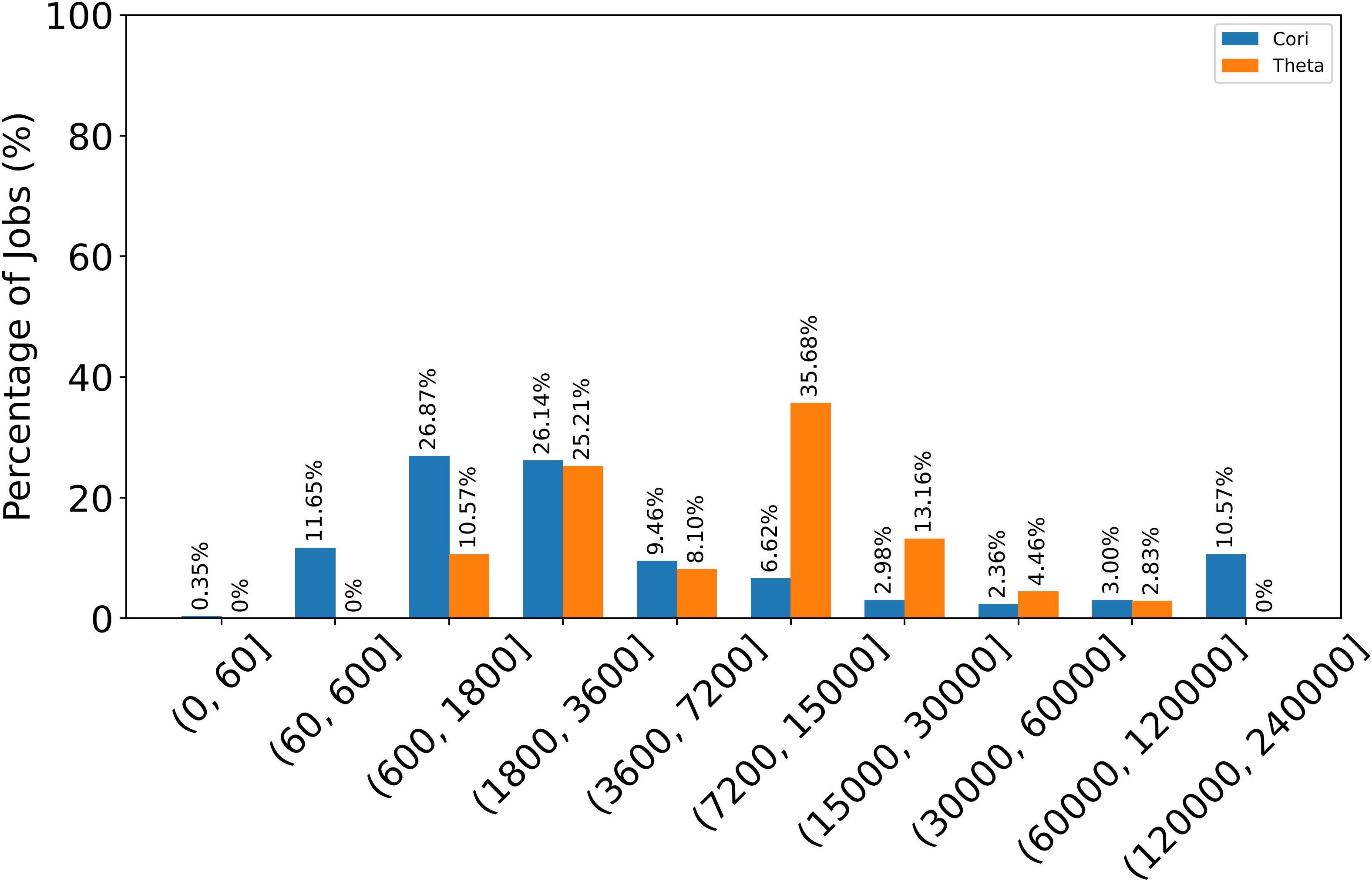}
        \label{fig:job_runtime_distribution}}
    \caption{\small{Job distribution of Theta and Cori workloads}}
    \label{fig:job_distribution}
    \vspace{-5pt}
\end{figure}

\begin{figure*} 
    \centering
    \subfloat[Theta]{%
       \includegraphics[width=0.49\linewidth]{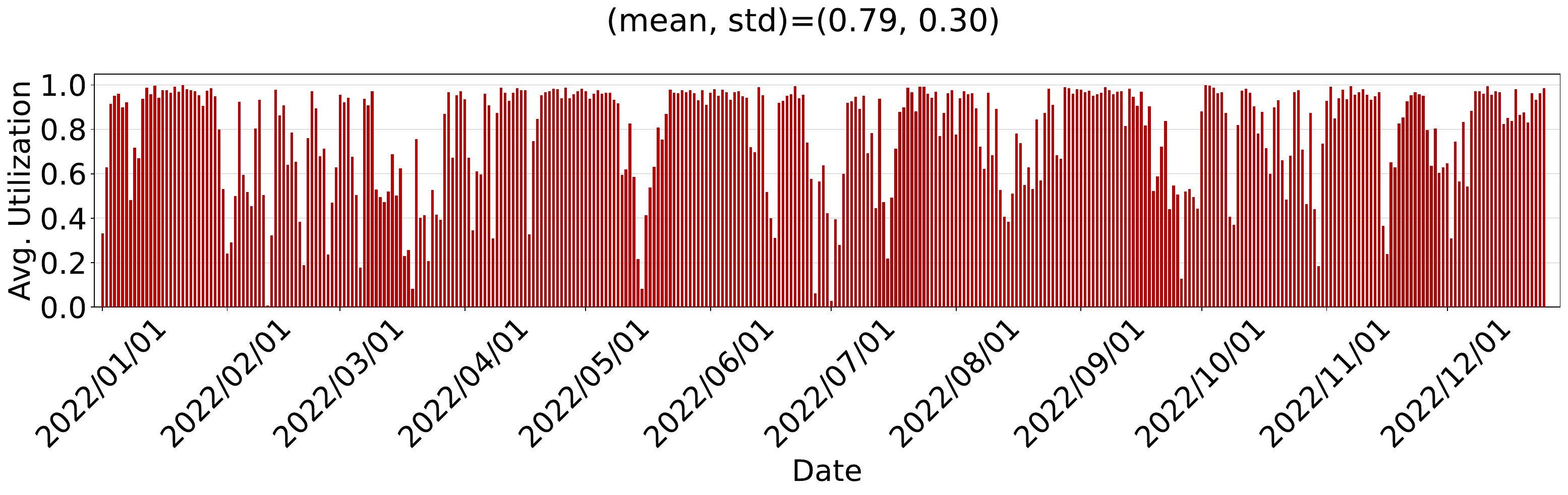}
       \label{fig:theta_utl}}\hfill
    \subfloat[Cori]{%
        \includegraphics[width=0.49\linewidth]{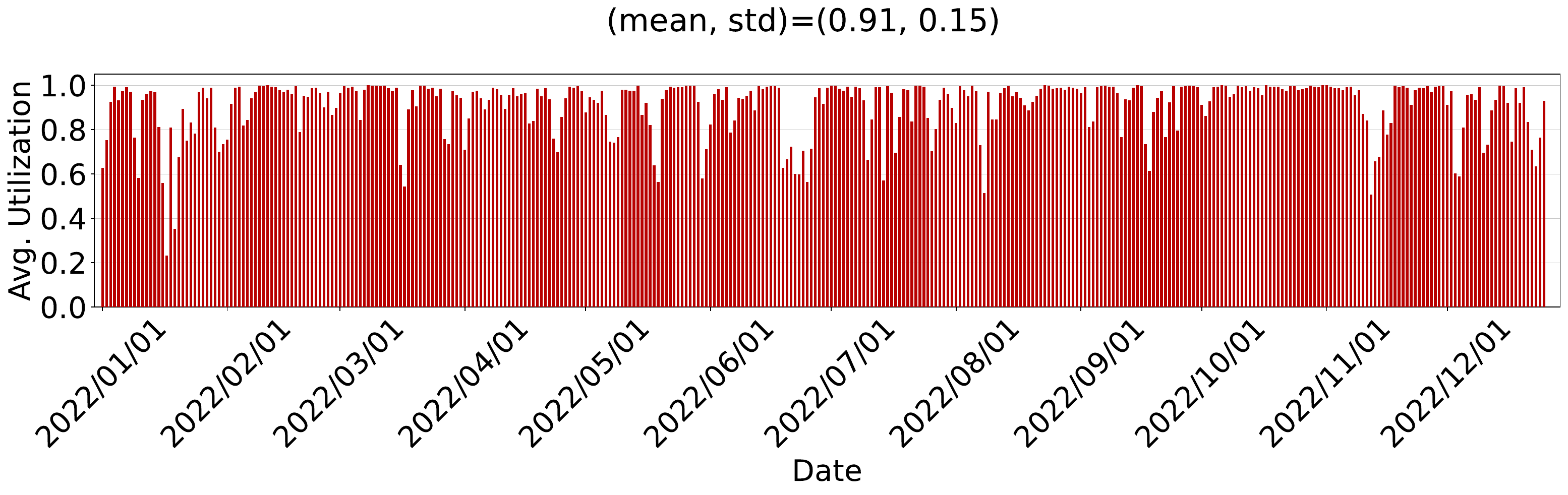}
        \label{fig:cori_utl}}
    \caption{Average daily utilization of Theta and Cori in the year 2022.}
    \label{fig:theta_cori_utl_2022}
\end{figure*}

Figure~\ref{fig:theta_cori_utl_2022} presents daily resource utilization on Theta and Cori. 
The utilization of resources can fluctuate significantly, experiencing periods of both high and low demand over days. This dynamic nature is influenced by various factors, such as the nature of computational workloads, user demands, and the specific applications running on the system. Finding ways to manage the burstiness of jobs in these systems is important. 

The resource utilization on Theta appears to be lower than that on Cori. 
Our detailed analysis indicates that this is due to Theta's spatial and temporal scheduling holes, as the machine is provisioned for large-sized and long-running jobs. 
Furthermore, Theta's utilization exhibits greater variance than Cori, suggesting significant fluctuations in resource demands on Theta.
Interestingly, high usage on one system may coincide with low usage on the other, suggesting an opportunity for workload balancing. For instance, while only eight jobs were submitted to Theta on 2/11/2022, Cori's utilization was 99.95\% on that day.


\section{Integration Methodology}
\label{method}
Motivated by the workload characterization, we argue that through \emph{strategic consolidation}, not only can we achieve better overall system utilization by filling each other's resource gaps, but we also have the opportunity to reduce redundant operational expenses (i.e., \emph{``more for less''}). Specifically, we conduct two types of  unification analysis.

\textbf{Workload Fusion.} 
A unified system is created by integrating capability-predominant and capacity-predominant computing systems. Simultaneously, a consolidated workload, encompassing both capability and capacity jobs, is executed on the unified system. Specifically, we examine the performance of the consolidated jobs on a unified system of 14,048 nodes ($4,360+9,688$).

Moreover, we downsize the unified system to 13,345 (95\%), 12,643 (90\%), 11,940 (85\%), and 11,238 (80\%) nodes, while keeping the workloads unchanged. Consequently, the workload fusion analysis provides insights into whether consolidating workloads can efficiently utilize the unified platform while also offering additional benefits, such as cost savings. Specifically, workload fusion aims to answer two fundamental questions: 

\begin{itemize}
\setlength{\parskip}{0pt}
\setlength{\itemsep}{0pt plus 1pt}
\item \emph{Q1. What impacts could arise from accommodating both capability-predominant and capacity-predominant computing workloads within a unified platform?}
\item \emph{Q2. To what extent can the size of the unified system be reduced to lower total ownership costs without compromising computational capabilities for a mix of jobs?
}
\end{itemize}

\textbf{Workload Injection.} The analysis aims to assess how effectively the scheduling gaps in a capability-predominant system (Figure~\ref{fig:scheduling hole}) can be filled by leveraging capacity jobs. We explore how to strategically select Cori jobs and analyze the impact of injecting Cori capacity jobs on Theta's capability computing jobs.

 
In this study, Cori jobs, submitted based on their original arrival times, 
are treated as backfilled jobs on Theta. We maintain two job queues: the default queue, which runs the Theta jobs, and the backfill queue, which receives Cori jobs according to their arrival times. On Theta, the minimum scheduling size is 128 compute nodes, meaning the scheduler allocates at least 128 nodes to a user job, even if the job size is smaller. The extra allocated nodes are wasted. Additionally, Theta experiences substantial spatial and temporal scheduling gaps due to the lack of small or short-running jobs in the capability workload. Workload injection seeks to utilize Cori’s capacity jobs to harness the idle resources on Theta. Specifically,  workload injection seeks to answer two fundamental questions: 

\begin{itemize}
\setlength{\parskip}{0pt}
\setlength{\itemsep}{0pt plus 1pt}
\item \emph{Q3. What are the potential impacts of accommodating additional capacity jobs on a capability-predominant computing system?}
\item \emph{Q4. How can we strategically select capacity jobs without compromising the performance of critical capability computing jobs?}
\end{itemize}

\textbf{Analysis Strategy.} 
We extend the open-source scheduling simulator named CQSim \cite{CQSim} for the integration analysis. CQSim originated from a discrete-event-driven scheduling simulator developed for the batch scheduler Cobalt at ALCF. The simulator has been extensively validated with real system traces listed in the well-known Parallel Workload Archive\cite{SWF,WFP_tang_2011,yang-sc13,allcock2017experience}. 
CQSim mimics the real-world job scheduling environment: in a real HPC system, the scheduler receives user job submissions during runtime; CQSim reads the job submissions from workload traces. We extend CQSim by implementing separate job queues to handle different types of jobs. This improved tool allows us to explore various design alternatives, such as different workload combinations on different system configurations. 

For workload fusion, we simulate a unified system composed of 14,048 Intel KNL nodes ($4,360+9,688$). We combine the two workloads based on job arrival times and feed the combined workload into the unified system. Additionally, we conduct extensive what-if analyses by reducing the system size and investigating how this impacts both capability and capacity jobs. 
To ensure fairness for different jobs on the unified system, we use the widely adopted FCFS scheduling with EASY backfilling for the workload fusion analysis.

For workload injection, we select various Cori sub-workloads, including small-sized jobs and short-running jobs, to investigate how the injection of these capacity jobs would impact Theta's workload. Cori sub-workloads are submitted to the backfill queue based on their arrival times, and these workloads are allocated for execution when idle resources are available on Theta, following the backfill policy.  We use the capability scheduling policy WFP (described in Section \ref{sec:2}) with EASY backfill for the workload injection analysis.

\textbf{Metrics.} We use the following scheduling metrics:

\begin{itemize}
  \setlength{\parskip}{0pt}
  \setlength{\itemsep}{0pt plus 1pt}
\item \textbf{Resource utilization} is a system-centric scheduling metric that measures the ratio of used compute nodes to the total number of compute nodes in the system. It measures how well the computing resource is utilized.  
\item \textbf{Job wait time} is a user-centric scheduling metric that measures the elapsed time between job submission and job start time. This metric directly measures the satisfaction of users. 
\item \textbf{Rescued resource} is defined as the percentage of wasted node-hours in a capability system utilized by additional workloads. 
This metric measures how scheduling gaps can be filled by incorporating additional capacity jobs.
\end{itemize}

\section{Workload Fusion}\label{fusion}

\begin{figure} 
    \centering
        \includegraphics[width=0.65\linewidth]{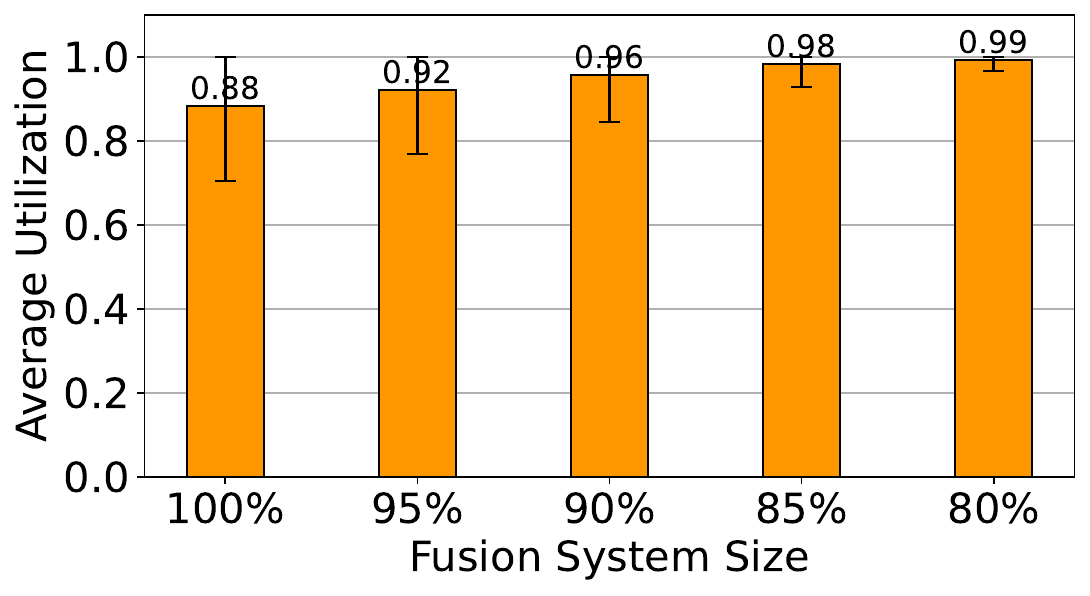}
    \caption{Resource utilization (mean and standard deviation) under workload fusion. The system size is 14,048 (100\%), 13,345(95\%), 12,643(90\%), 11,940(85\%), and 11,238(80\%).}
    \label{fig:fusion_utl}
    \vspace{-5pt}
\end{figure}

\begin{figure} 
    \centering
    \subfloat[]{%
       \includegraphics[width=0.48\linewidth]{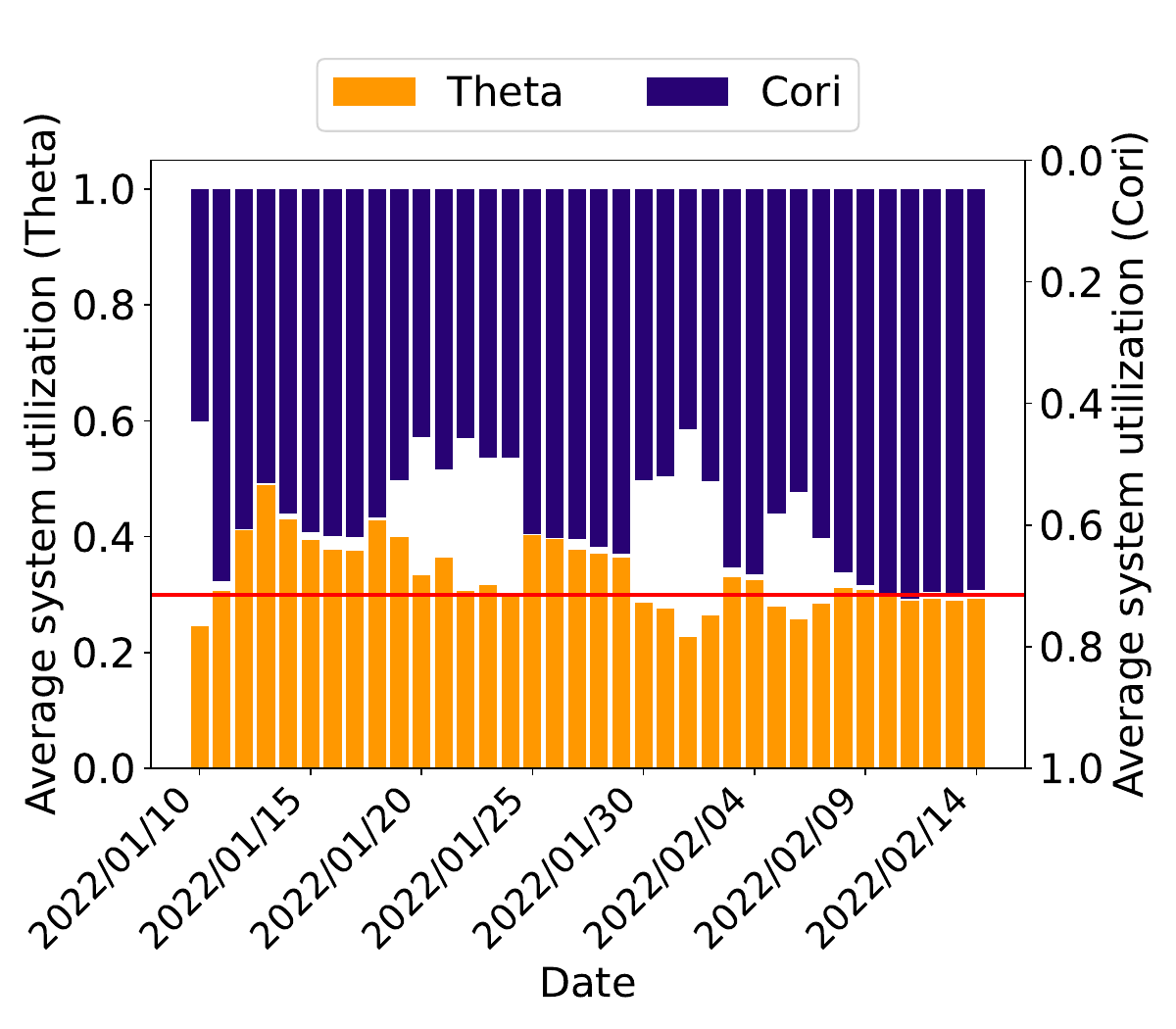}
       \label{fig:fusion_theta_utilize_cori}}\hfill
    \subfloat[]{%
        \includegraphics[width=0.48\linewidth]{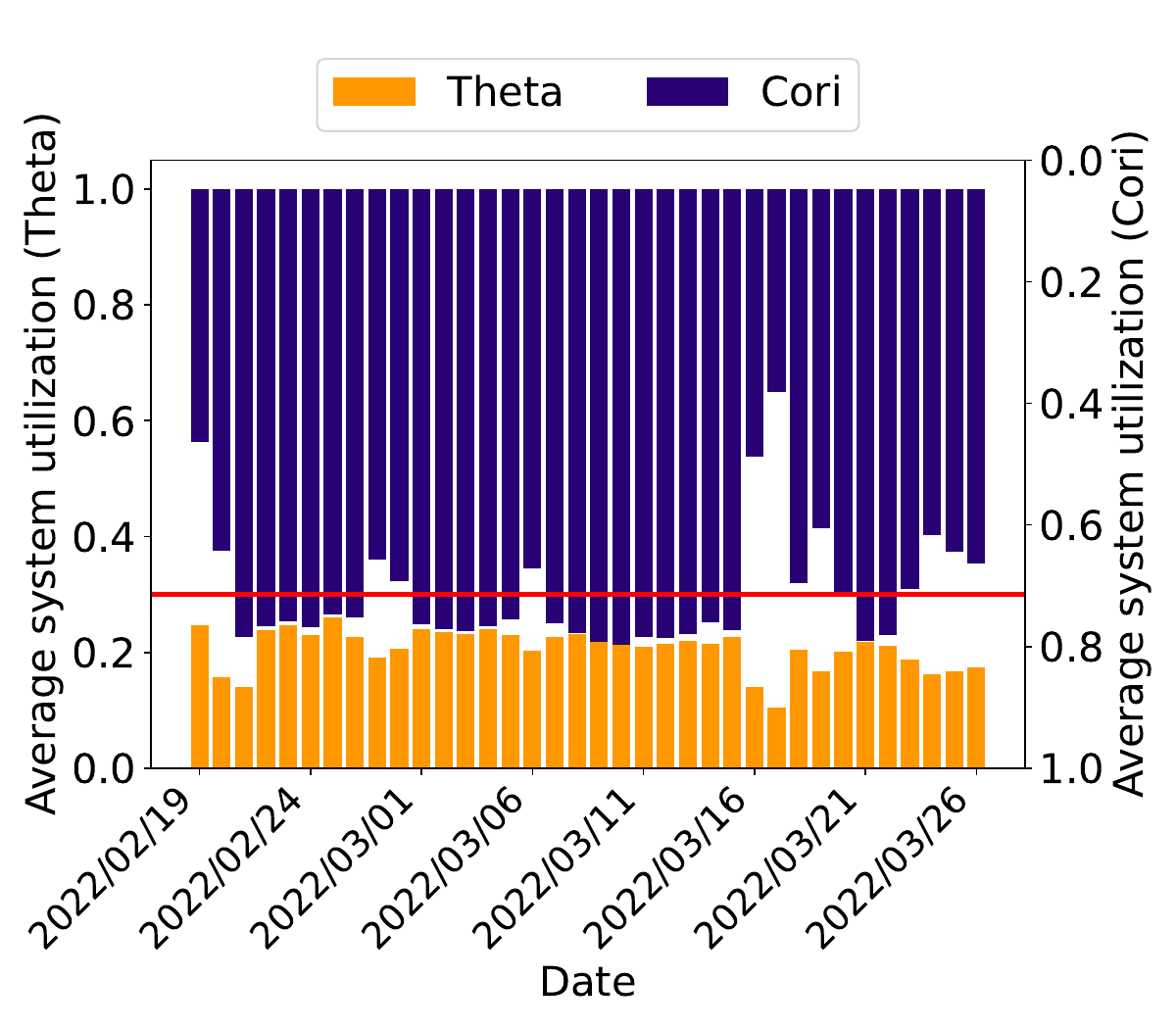}
        \label{fig:fusion_cori_utilize_theta}}
    \caption{Theta utilization is shown from bottom to top, and Cori utilization is shown from top to bottom. The red horizontal line marks the boundary between the two systems, with Theta's resources contributing approximately 31\% to the unified system.
    (a) Theta jobs benefit from the unified system, and (b) Cori jobs benefit from the unified system.
    }
    \label{fig:fusion_theta_cori_utilize_each_other}
    \vspace{-5pt}
\end{figure}

Figure \ref{fig:fusion_utl} shows the average resource utilization and variation under workload fusion with varying sizes of the unified system. Here, ``100\%'' represents the original unified system composed of $14,048$ nodes, while ``90\%'' represents a unified system with 10\% size reduction. As we gradually reduce the system size, we observe an increase in system utilization of up to 12\%, reaching as high as 99.6\% when the size of the unified system is reduced by 20\%. Moreover, we observe a decrease in the variance of system utilization as the system size decreases, suggesting that the system becomes saturated.

When co-scheduling the workloads on a unified system, the idle resources from Theta and Cori can be utilized by diverse jobs in the unified system. To gain a deep understanding of this phenomenon, we present two case studies for a unified system at full scale (100\%) in Figure \ref{fig:fusion_theta_cori_utilize_each_other}:

\begin{itemize}
  \setlength{\parskip}{0pt}
  \setlength{\itemsep}{0pt plus 1pt}
\item 
Theta reached its peak capacity from 01/11/2022 to 01/29/2022, while Cori coincides with low utilization during that period (Figure \ref{fig:fusion_theta_utilize_cori}). We can see that Theta jobs contribute more than 31.2\% to the utilization of the unified system on some days between 2022/01/11 and 2022/01/19, indicating that the Theta jobs are utilizing the idle resource of the Cori system.
\item 
Cori experiences high demands of resources during 02/19/2022 - 03/26/2022, while Theta coincides with low utilization during that same period (Figure \ref{fig:theta_cori_utl_2022}). 
We observe that, during this period, Cori jobs contributed more than 68.8\% to the utilization of the unified system. This indicates that during this period, Cori's jobs are effectively utilizing the idle resources of the Theta system.
\end{itemize}

\begin{mybox}
\textbf{Observation 4.}  The fusion of systems and workloads provides opportunities for load balancing, enabling mixed workloads to better utilize the integrated computing resources for different tasks, ultimately leading to greater efficiency in the unified system.
\end{mybox}

\begin{mybox}
\textbf{Observation 5.} 
By consolidating various jobs within a unified system, we can significantly reduce operational expenses associated with infrastructure maintenance and software management. Additionally, the unified system can be downsized for further cost savings while achieving higher resource utilization --- more for less.
\end{mybox}

\begin{figure} 
    \centering
    \centering
    \subfloat[Average Theta job wait time]{%
       \includegraphics[width=0.48\linewidth]{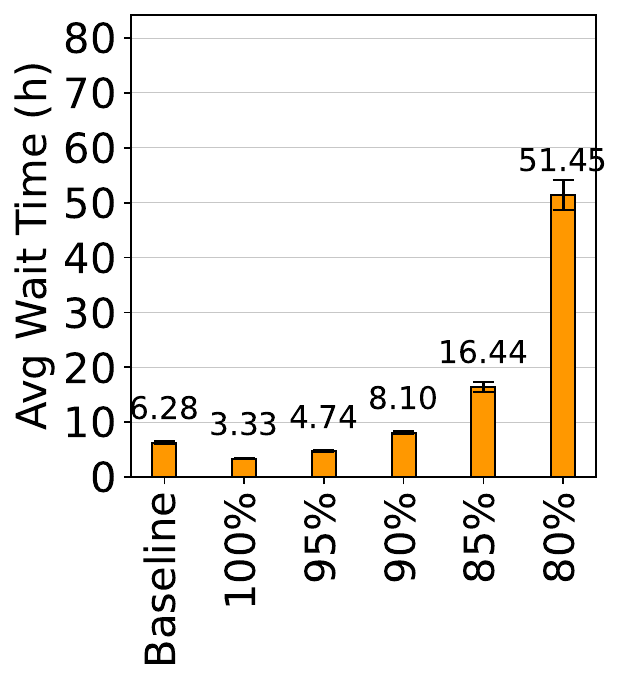}
       \label{fig:fusion_theta_avg_wait}}\hfill
    \subfloat[Average Cori job wait time]{%
        \includegraphics[width=0.48\linewidth]{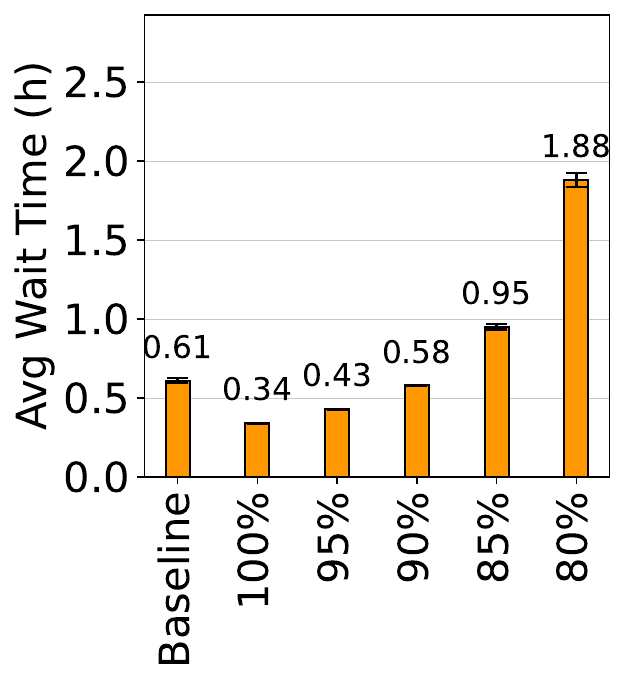}
        \label{fig:fusion_cori_avg_wait}}\hfill
    \caption{Average job wait time under workload fusion. These plots show the standard error with a 95\% confidence interval. This indicates that we are 95\% confident the job wait times fall within the confidence interval for each case.}
    \vspace{-5pt}
\end{figure}

Next, we evaluate the impact of system size reduction on job wait times in the unified system.
Figure \ref{fig:fusion_theta_avg_wait} and \ref{fig:fusion_cori_avg_wait} show the effect of workload fusion on job wait times under varying system sizes. In the unified system(100\%), the average job wait time is significantly reduced  compared to the silo case (i.e., the baseline). The average wait time for Theta jobs is reduced from 6.28 to 3.3 hours (1.9X). Similarly, the average wait time for Cori jobs is reduced from 0.6 to 0.31 hours (1.9X). This reduction occurs because capability-predominant and capacity-predominant workloads can utilize a larger pool of resources in the unified system, thereby reducing the average job wait time for both types of jobs.

\begin{figure*} 
    \centering
    \subfloat[Theta workload by different job sizes]{%
       \includegraphics[width=0.48\linewidth]{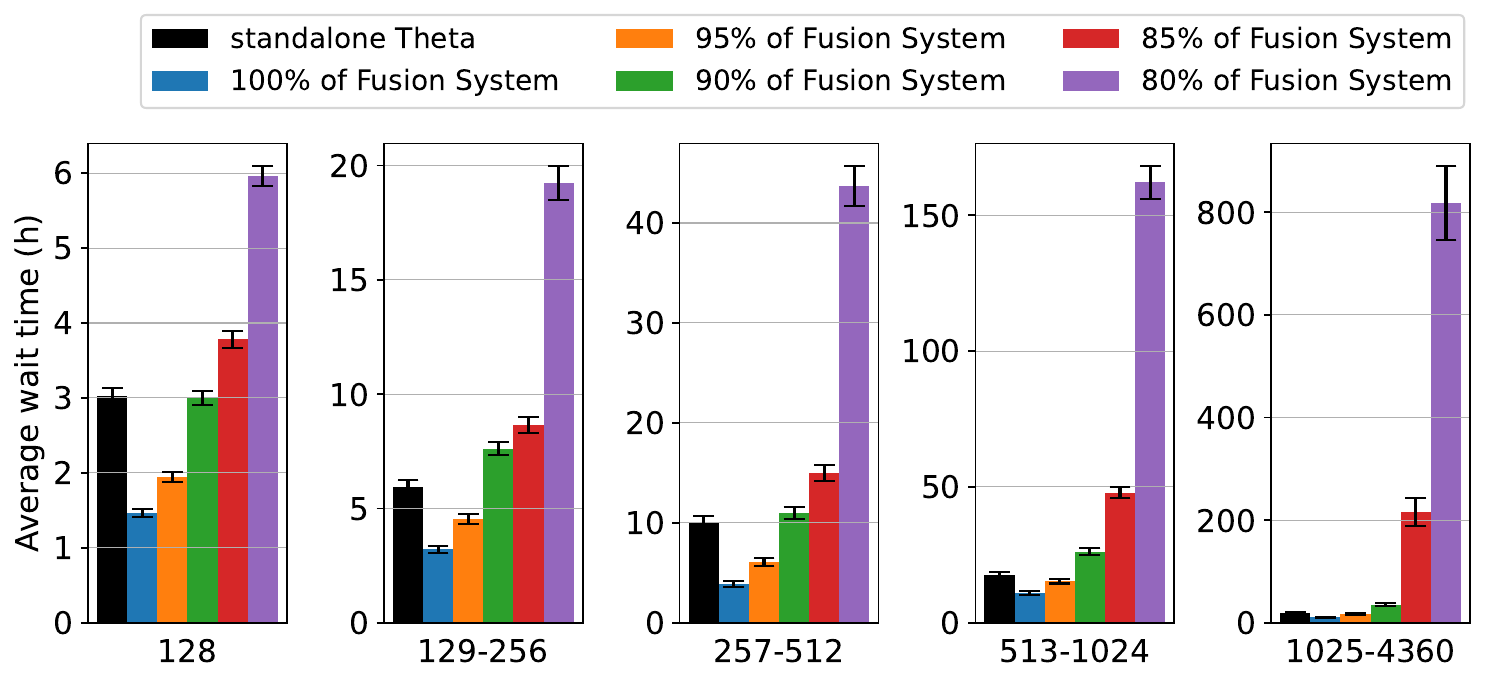}
       \label{fig:fusion_theta_avg_wait_diff_size}}\hfill
    \subfloat[Theta workload by different job runtimes]{%
        \includegraphics[width=0.48\linewidth]{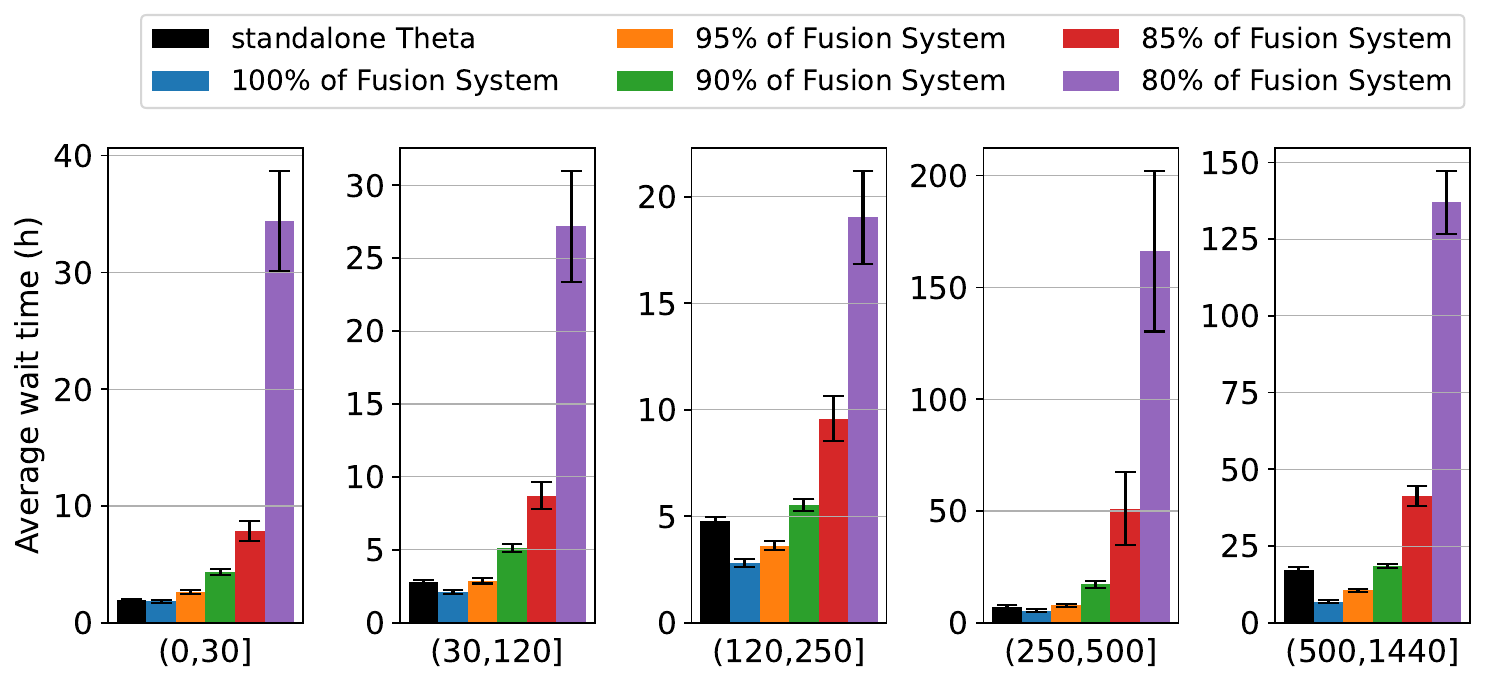}
        \label{fig:fusion_theta_avg_wait_diff_runtime}}

    \subfloat[Cori workload by different job sizes]{%
       \includegraphics[width=0.48\linewidth]{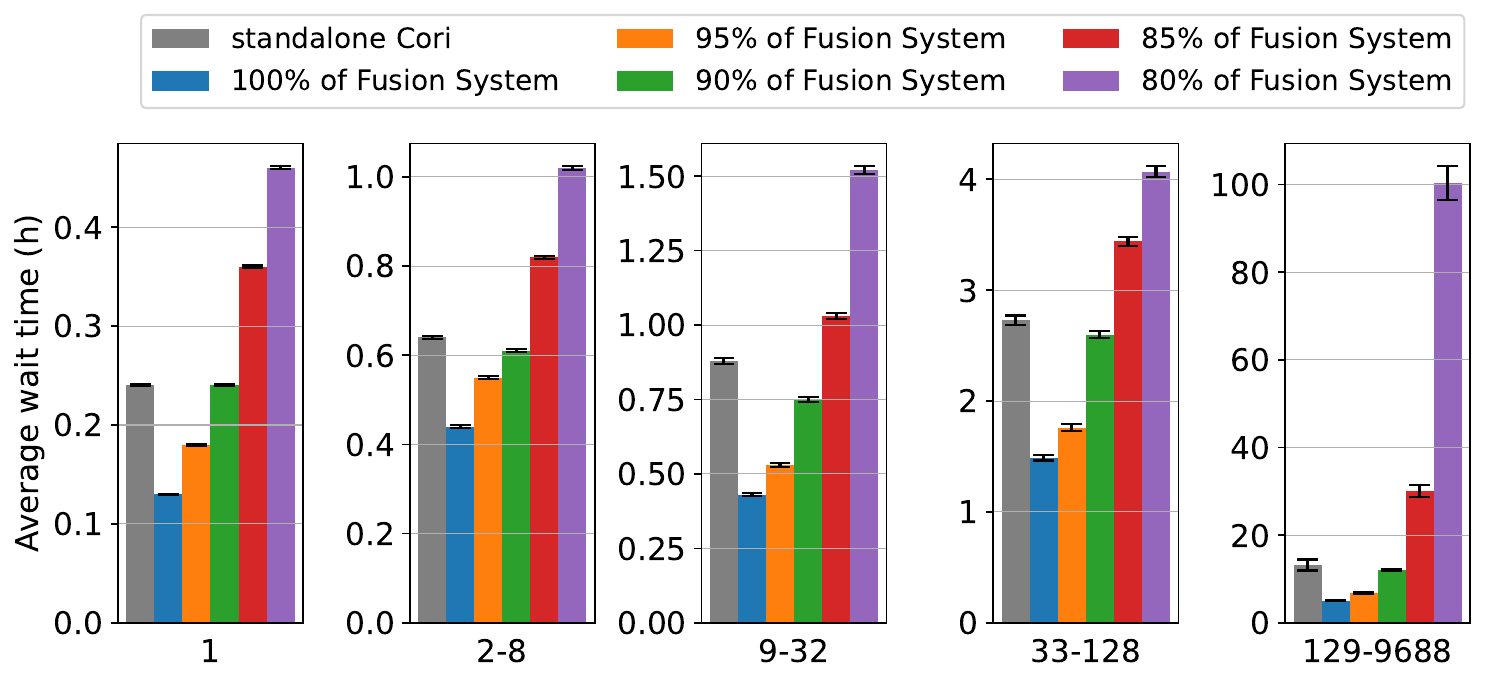}
       \label{fig:fusion_cori_avg_wait_diff_size}}\hfill
    \subfloat[Cori workload by different job runtimes]{%
        \includegraphics[width=0.48\linewidth]{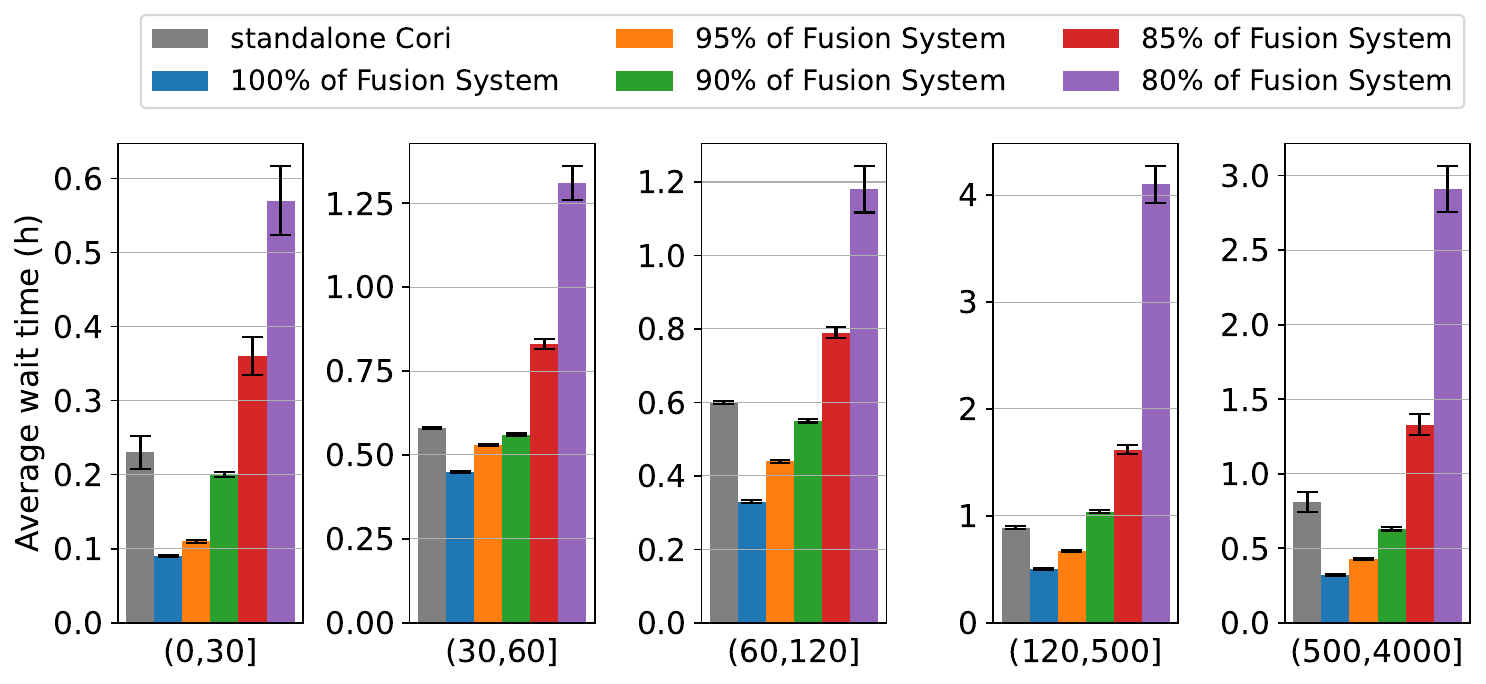}
        \label{fig:fusion_cori_avg_wait_diff_runtime}}\hfill
       \caption{Average job wait times of different job types in the unified system. The standard error with a 95\% confidence interval is shown, indicating that we are 95\% confident the job wait times fall within the interval for each case.}
    \label{fig:fusion_avg_wait_time_by_category} 
    \vspace{-5pt}
\end{figure*}

When we reduce the unified system size to 90\%, there is no adverse impact on job wait time for either Theta or Cori workloads compared to the baselines. This is encouraging in terms of the cost-effectiveness of system integration. 
According to \cite{rountree2016}, the cost of a rack of 100 server nodes is about \$366,000. 
A 10\% system reduction (1,400 nodes) could result in at least \$5 million ($14\times \$366,000$) in savings for machine procurement costs.

However, we observe a significant increase in job wait time for both Theta and Cori when the unified system size is reduced to 80\%. 
Hence, capacity planning is crucial for making informed decisions on reducing system size.

\begin{mybox}
\textbf{Observation 6.} Integrating a variety of jobs within a unified system significantly reduces job wait time --- up to twice as fast --- leading to a substantial improvement in user experience compared to the isolated scenario.
\end{mybox}

\begin{mybox}
\textbf{Observation 7.} User experience, represented by job wait time, remains unaffected when the unified system is downsized by up to 10\%. However, a more pronounced impact on job wait time is observed when the unified system is downsized by 20\%.
\end{mybox}

We also delve deeper into the impacts on different categories of jobs, such as job size and runtime, shown in Figure \ref{fig:fusion_avg_wait_time_by_category} (a)-(d).
As the size of the unified system is reduced, both Theta and Cori jobs across all categories experience longer wait times. For both workloads, smaller jobs are less sensitive to changes in the size of the unified system, while larger jobs are more affected. For instance, the average wait time for Theta jobs requesting 128 nodes is approximately twice as long in a system reduced to 80\% of its capacity compared to the full 100\% capacity. 
However, the average wait time for jobs requesting more than 1,024 nodes is around 40 times longer! This is expected, as reducing the size of the unified system lowers the likelihood that extremely large jobs would be scheduled quickly, given the reduced available computing capacity, while smaller jobs can still be backfilled. 
Similarly, Cori jobs requesting no more than 128 nodes experience a wait time twice as long when the unified system size is reduced to 80\%. However, for Cori jobs requesting more than 128 nodes, the wait time increases by 7.6 times.

When reducing the unified system size, Theta jobs with longer job runtimes tend to have much longer wait times than compared to shorter jobs. For example, Theta jobs, requesting more than 250 minutes, experience more than 100 minutes of wait time, while the average wait time is no more than 30 minutes for Theta jobs requesting 0 to 250 minutes of runtime. This is because longer Theta jobs are typically larger, as shown in Figure \ref{fig:theta_dist}. 
 As a result, when the unified system shrinks, these longer and larger jobs are more likely to experience delays. Similarly, Cori jobs requesting more than 120 minutes of runtime also face longer wait times than other jobs.

\begin{mybox}
\textbf{Observation 8.} Jobs of varying sizes, whether capability or capacity, exhibit different degrees of sensitivity to changes in the size of the unified system. Typically, smaller jobs are less sensitive to changes in system size, while larger jobs are more affected by such modifications.
\end{mybox}

\begin{mybox}
\textbf{Observation 9.} 
Long-running jobs, whether capability or capacity, are often large. Therefore, when the unified system is reduced by more than 10\%, these jobs tend to experience significantly longer wait times compared to shorter jobs.
\end{mybox}

This workload fusion analysis addresses \emph{Q1 and Q2} listed in Section~\ref{method}. 
First, workload fusion within a unified system offers several advantages, including improved workload balancing, more efficient resource utilization, and reduced job wait times. 
Additionally, by consolidating a variety of jobs on a single platform, operational expenses such as infrastructure maintenance and software management can be reduced, leading to cost savings. 
Second, a moderate reduction in the size of the unified system, typically up to 10\%, can be achieved without compromising computational capabilities for various jobs with diverse computing requirements.

\section{Workload Injection}\label{injection}

As shown in Figure \ref{fig:theta_cori_utl_2022}, the capability-predominant system, Theta, has lower utilization with higher variance than Cori. 
Here we examine how to effectively boost the resource utilization of the capability system by leveraging Cori jobs. Specifically, we conduct two injection case studies:

\begin{enumerate}
    \item Injection of \emph{small-sized jobs.} Given that the minimum job allocation size on Theta is 128 compute nodes and any unused allocation is wasted, we select Cori jobs with a size of 128 nodes or fewer and inject them onto Theta based on their arrival times.  
    \item Injection of \emph{short-running jobs.} 
    We select Cori jobs with runtimes of 30 minutes or less and inject them onto Theta based on their arrival times. The goal is to leverage these short-running Cori jobs to fill the temporal scheduling gaps on Theta. 
\end{enumerate}

For both case studies, the injected Cori jobs are treated as \emph{backfilled jobs} scheduled to fill spatial or temporal scheduling holes on the capability system Theta. 

\begin{table}
    \centering
    \caption{Injection of Cori jobs onto Theta. W1-W3 comprise small-sized Cori jobs, while W4-W6 consist of short-running Cori jobs. 
    Since Theta contains 4,360 nodes, we limit the selection of jobs in W4-W6 to 4,096 nodes.}
    \label{tab:injected_Cori}
    \begin{tabular}{|c|c|c|c|} \hline 
         & Job size (nodes) & Job runtime (min) & Job counts \\ \hline \hline
         W1 & $\leq 128$ & $\leq 30$ & 521,574 \\ \hline 
         W2 & $\leq 128$ & $\leq 45$ & 629,475 \\ \hline 
         W3 & $\leq 128$ & $\leq 60$ & 803,285 \\ \hline \hline
         W4 & $\leq 4096$ & $\leq 10$ & 133,914 \\ \hline
         W5 & $\leq 4096$ & $\leq 29$ & 268,959 \\ \hline
         W6 & $\leq 4096$ & $\leq 30$ & 528,243 \\ \hline\hline
    \end{tabular}
\end{table}

\begin{figure}
    \centering
    \includegraphics[width=0.65\linewidth]{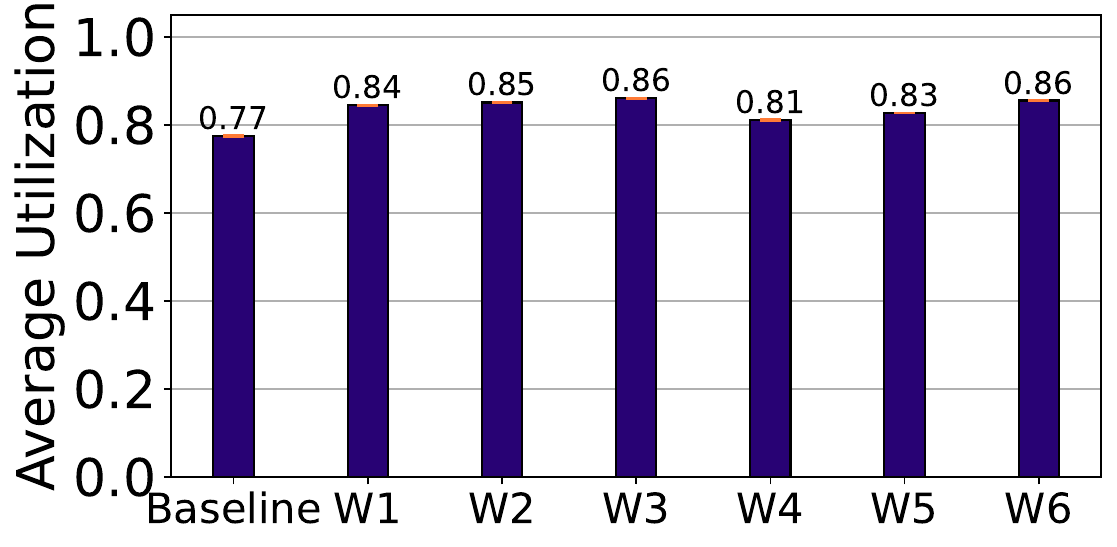}
    \caption{Theta utilization (mean and standard deviation) under different workload injection cases. The baseline is the case without workload injection.}
    \label{fig:injection_theta_avg_utl}
\end{figure}

Table \ref{tab:injected_Cori} lists six different Cori workloads used in the injection analysis. Workloads W1-W3 are used for \emph{the small-sized injection analysis}, whereas W4-W6 are used for \emph{the short-running injection analysis}. Workload W5 is chosen due to a significant difference in the number of jobs completed within 29 and 30 minutes, with the latter being twice as many. 

\begin{figure*} 
    \centering
    \subfloat[W1]{%
       \includegraphics[width=0.32\linewidth]{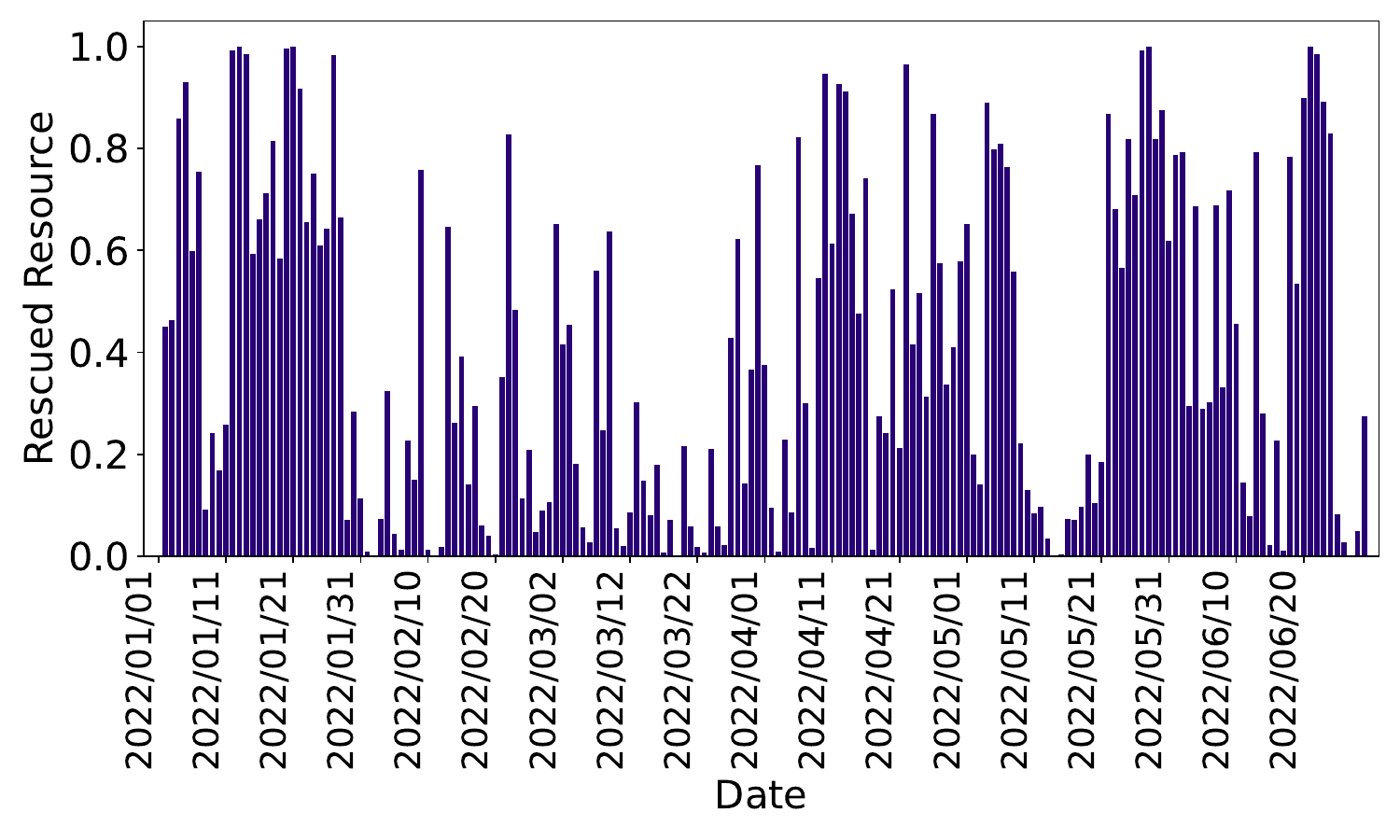}
       \label{fig:injection_rescued_w1}
    }
    \subfloat[W2]{%
        \includegraphics[width=0.32\linewidth]{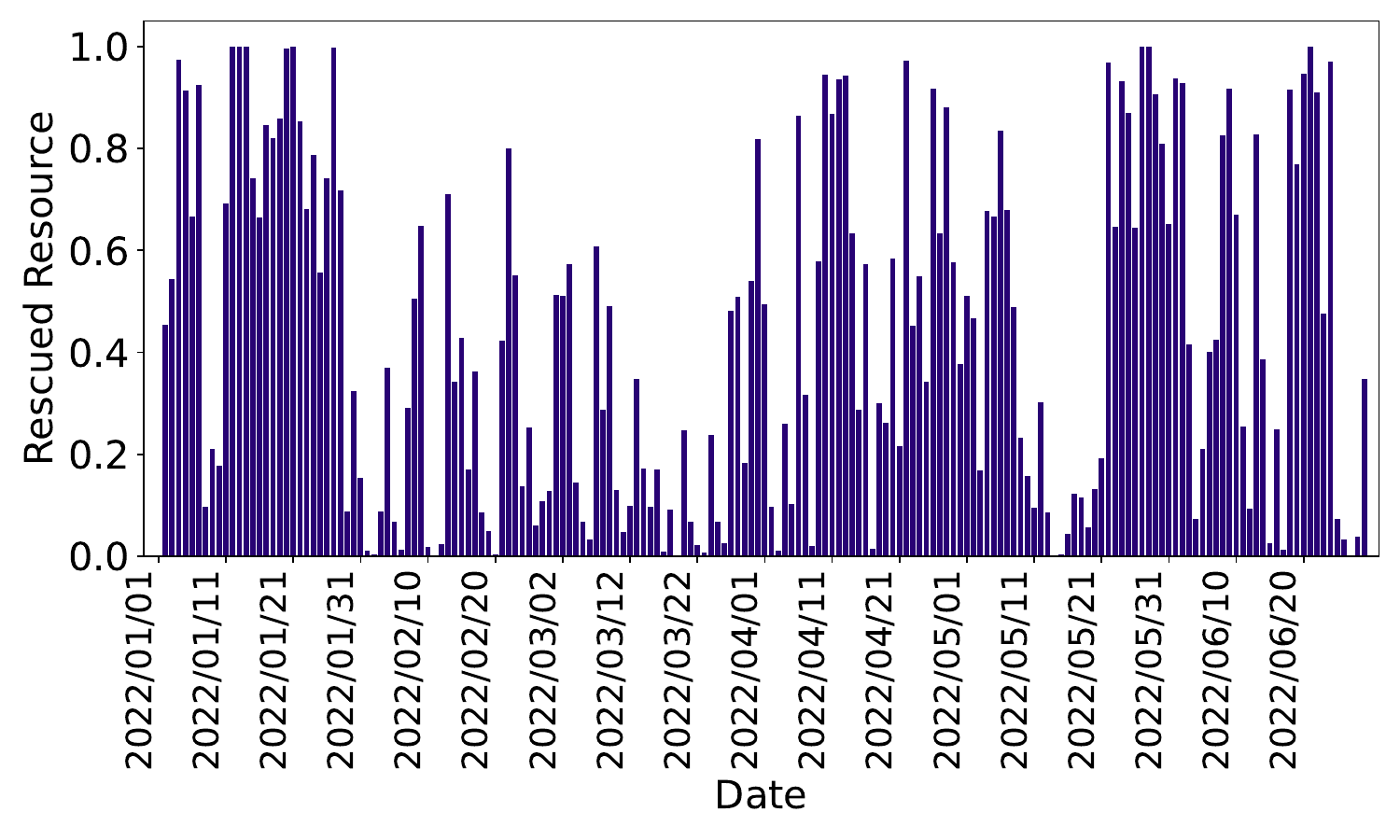}
        \label{injection_rescued_w2}
    }
    \subfloat[W3]{%
        \includegraphics[width=0.32\linewidth]{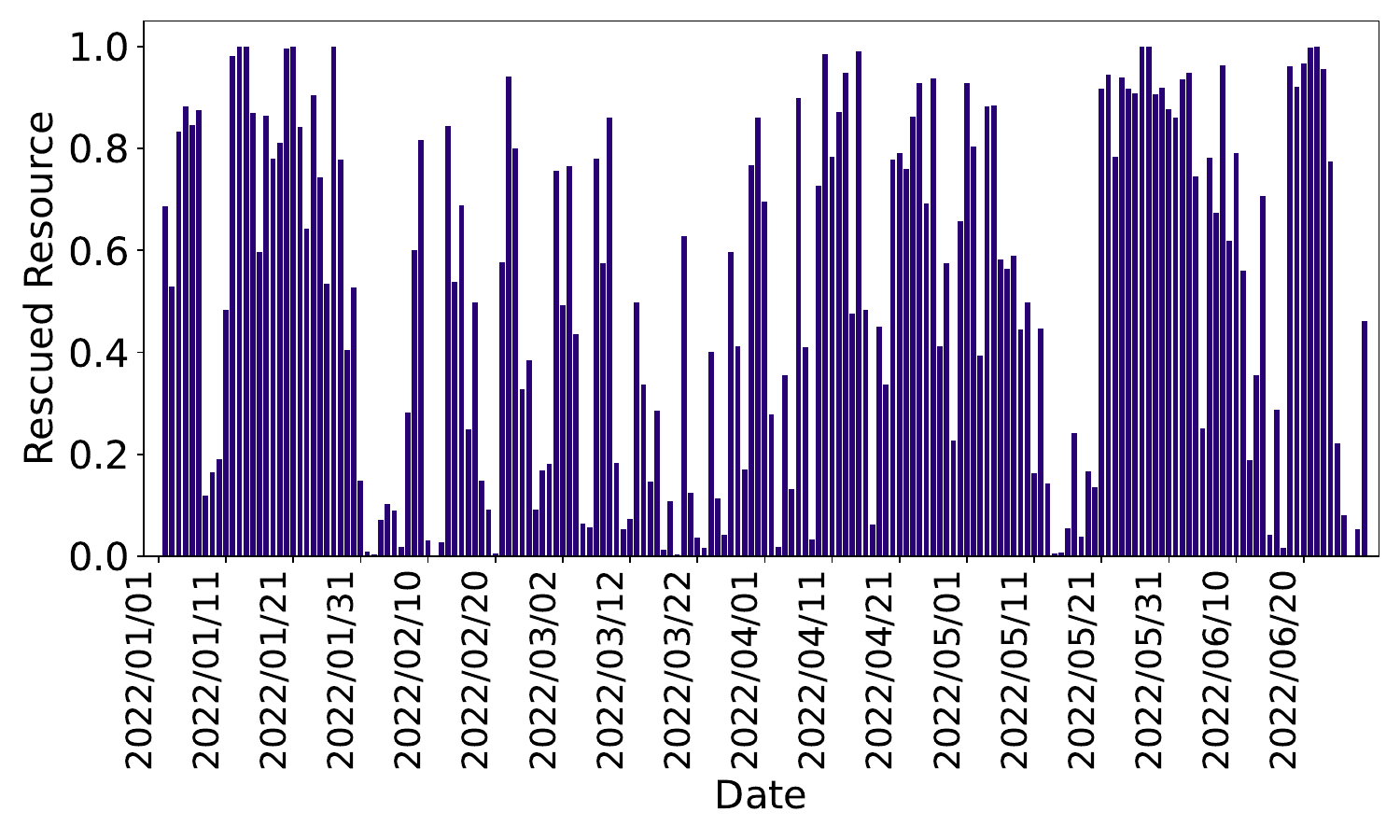}
        \label{injection_rescued_w3}
    }\\ 

    \subfloat[W4]{%
       \includegraphics[width=0.32\linewidth]{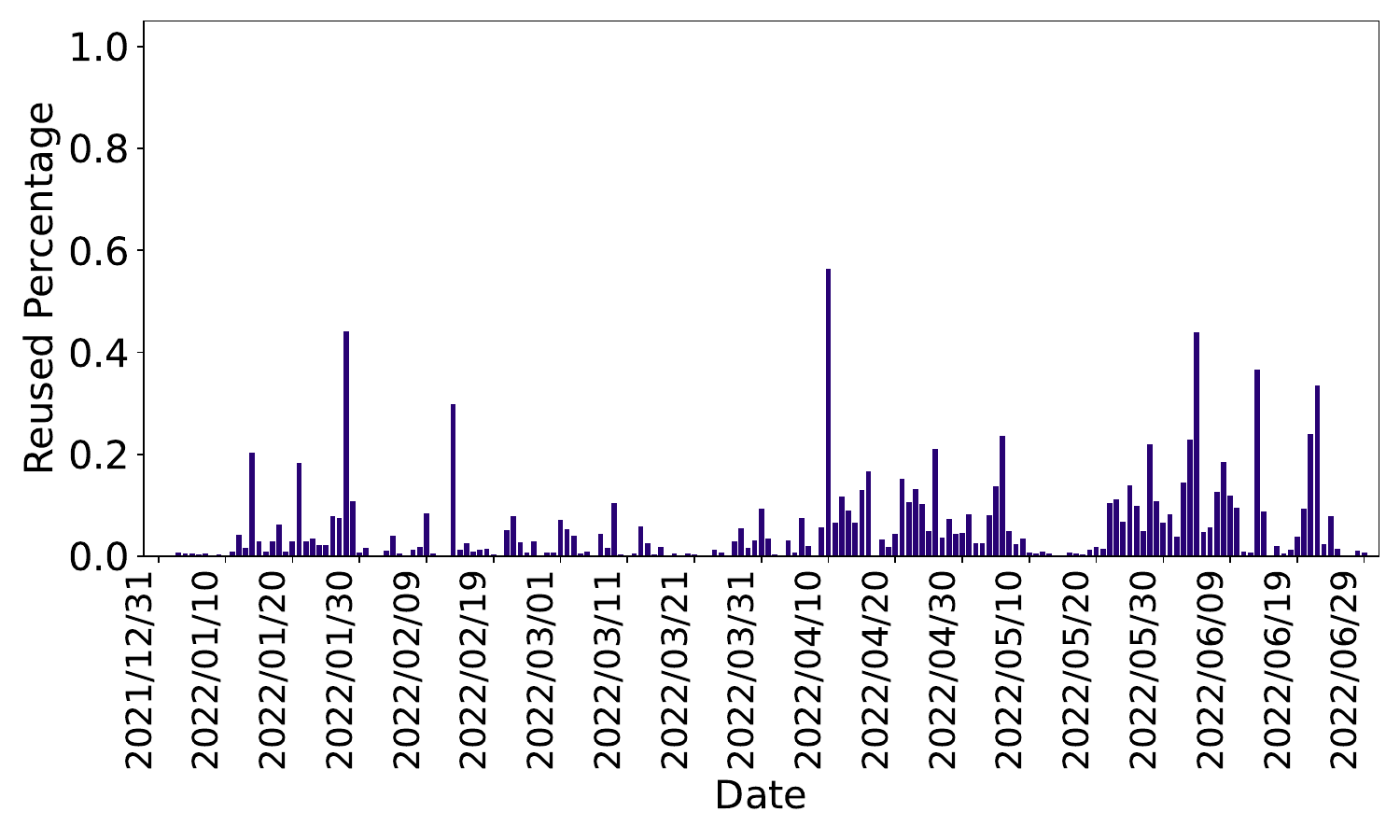}
       \label{injection_rescued_w4}
    }
    \subfloat[W5]{%
        \includegraphics[width=0.32\linewidth]{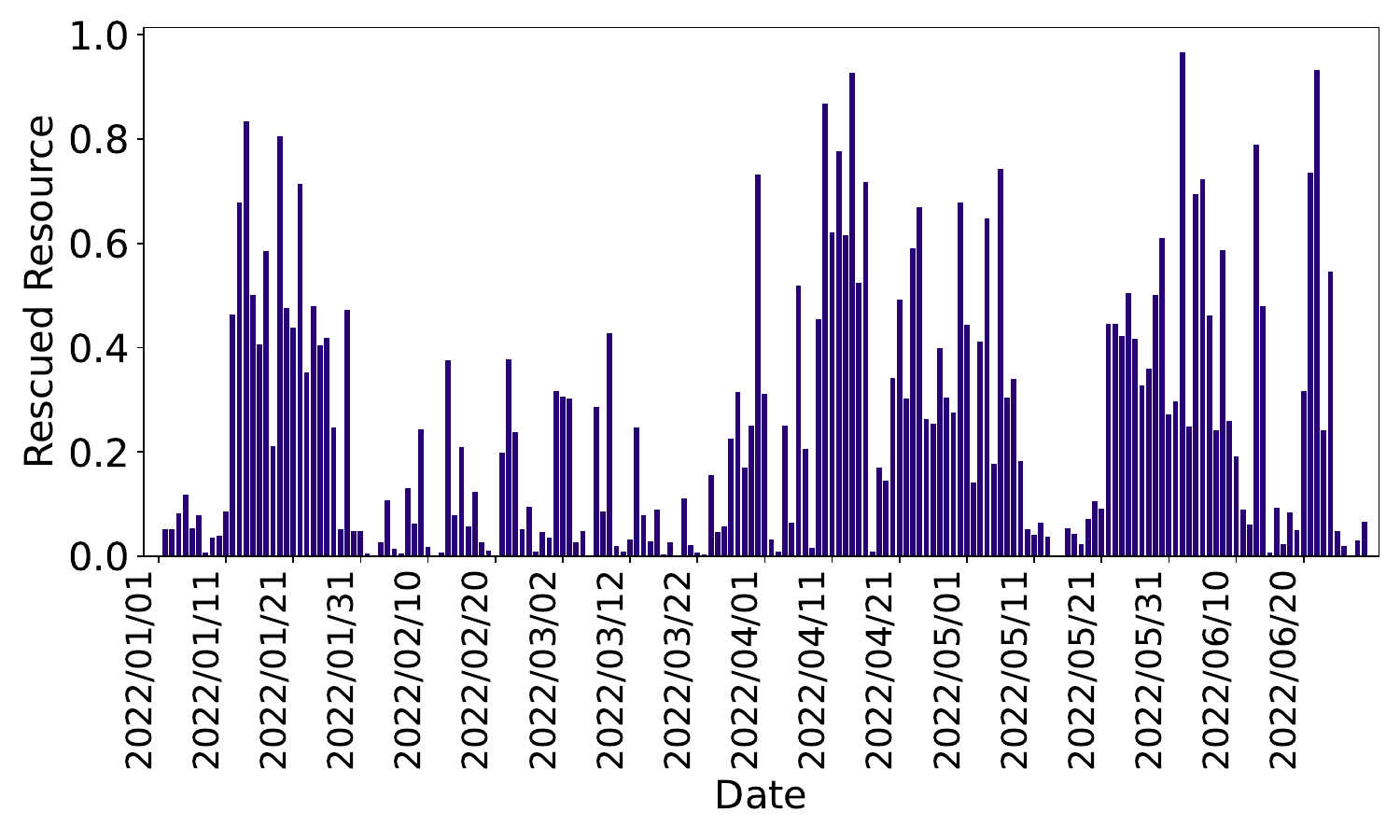}
        \label{injection_rescued_w5}
    }
    \subfloat[W6]{%
        \includegraphics[width=0.32\linewidth]{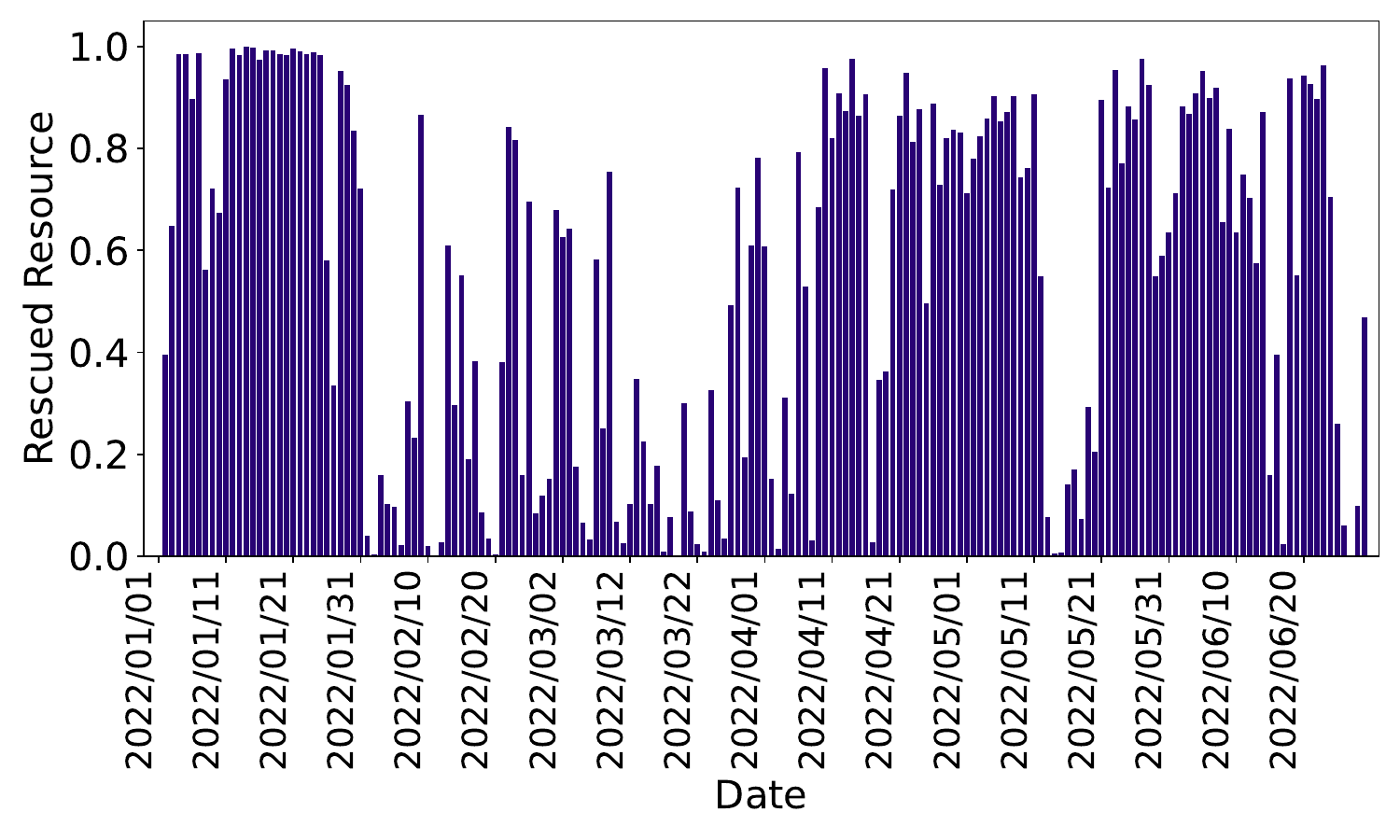}
        \label{injection_rescued_w6}
    }
    \caption{Rescued resource, percentage of wasted node-hours on Theta being utilized by capacity jobs, under workload injection. }
    \label{fig:injection_rescued} 
    \vspace{-0pt}
\end{figure*}

\begin{figure} 
    \centering
    \subfloat[]{%
       \includegraphics[width=0.65\linewidth]{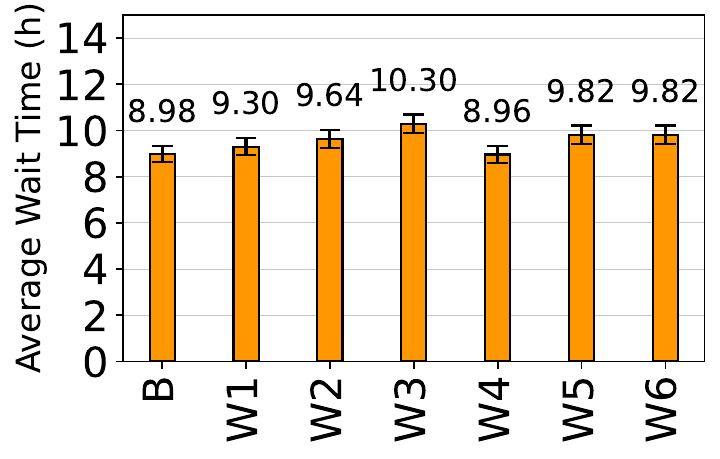}
       \label{fig:injection_theta_avg_wait_FCFS}}\hfill
    \caption{Job wait time of Theta workload under workload injection. "B" stands for baseline, which represents the case without workload injection. These plots display the standard error with a 95\% confidence interval. }
    \label{fig:injection_avg_wait}
    \vspace{-5pt}
\end{figure}
 
\begin{figure*} 
    \centering
    \subfloat[Job wait time under W1,W2,W3]{%
       \includegraphics[width=0.49\linewidth]{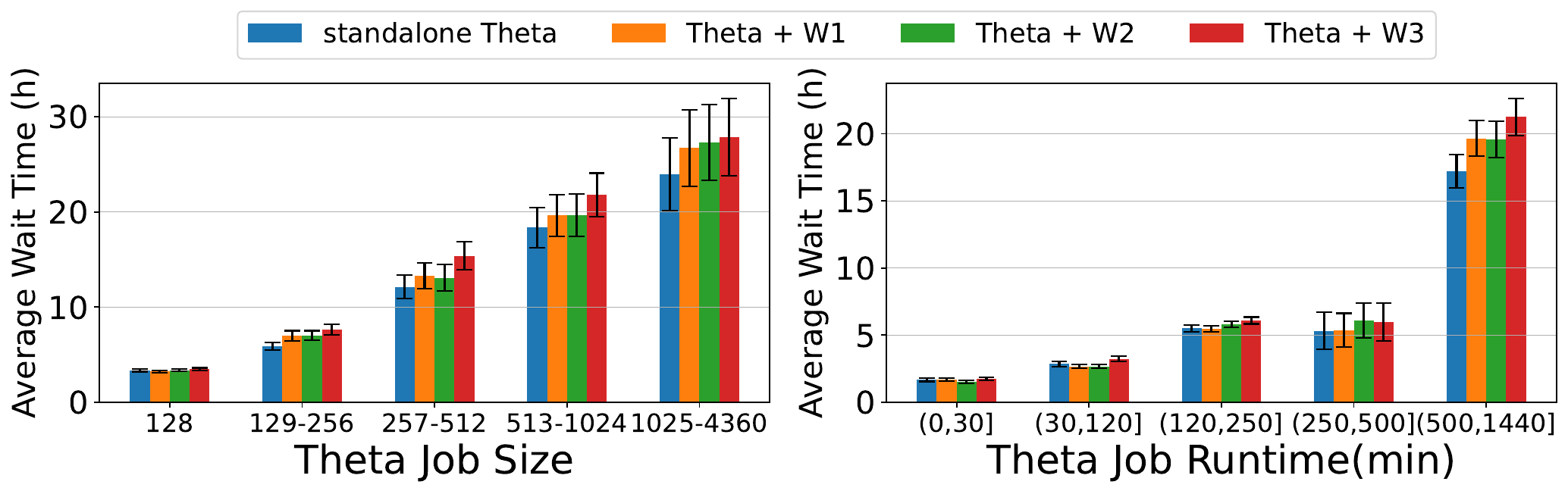}
      \label{fig:injection_theta_avg_wait_W123}}\hfill
    \subfloat[Job wait time under W4,W5,W6]{%
        \includegraphics[width=0.49\linewidth]{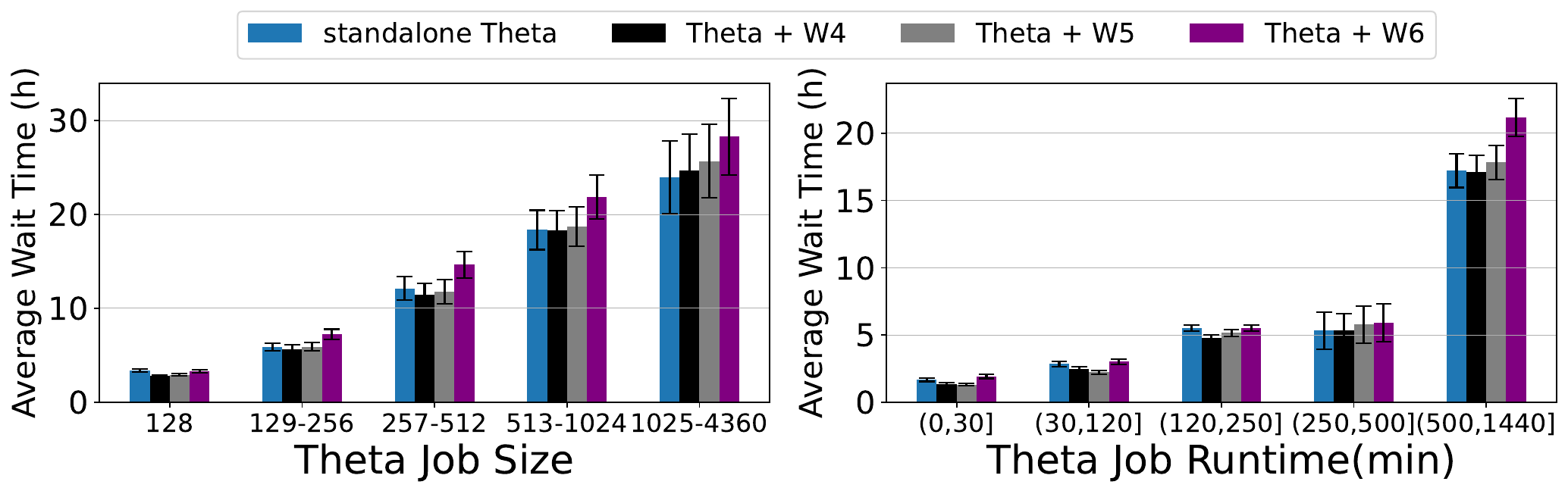}
    \label{fig:injection_theta_avg_wait_W456}}
    \caption{Job wait time of Theta workload categorized by job runtime and job size under workload injection. These plots show the standard error with a 95\% confidence interval. }
    \label{fig:injection_avg_wait_time_by_category}
    \vspace{-5pt}
\end{figure*}

Figure \ref{fig:injection_theta_avg_utl} presents the resource utilization of Theta after the injection of different workloads. The average resource utilization shows varying improvements under different injected Cori workloads.  For example, in the baseline scenario without injected Cori workload, the average utilization of Theta is approximately 77\%. When we inject the Cori workload W1, the average utilization  increases by 7\%, reaching 84\%. 
This demonstrates Theta's capability to utilize small-sized, short-running capacity jobs to address temporal and spatial allocation gaps, thereby enhancing resource efficiency.
From W1 to W3, where more small-sized jobs are injected, and from W4 to W6, where more short-running jobs are injected, we observe higher system utilization. This suggests that the capability system effectively utilizes small-sized or short-running capacity jobs to address its temporal and spatial allocation gaps.

Next, we study the impacts of workload injection in terms of \emph{rescued resource}, which directly measures the percentage of idle resources utilized by the newly injected capacity jobs on the capability system Theta. The results are shown in Figure~\ref{fig:injection_rescued}. 
Daily rescued resource ranges between zero and close to 100\%, influenced by allocation holes and the availability of capacity jobs.
It is interesting to note that in both cases of small-sized injection and short-running injection, Cori jobs can opportunistically utilize the idle resources on Theta, approaching 100\% on a number of days.
In the case of injecting small-sized jobs, we observe a gradual increase in the rescued resource metric from W1 to W3. Conversely, when injecting short-running jobs, we notice a drastic increase in the rescued resource metric from W4 to W6. It's worth noting that the number of short-running jobs increases by almost 4X from W4 to W6, while the number of small-sized jobs moderately increases by about 1.5X from W1 to W3.

Figure~\ref{fig:injection_avg_wait} depicts job wait times under various job injection scenarios. It is important to note that the injected Cori jobs are treated as backfill jobs. While backfill jobs do not directly affect the reserved Theta jobs, they may obstruct Theta jobs arriving after the backfill process. Consequently, Theta jobs experience extended wait times when additional capacity jobs are injected. We aim to quantify the extent to which workload injection contributes to this effect. It is observed that the average job wait time gradually increases with more capacity jobs injected, both in cases of small-sized and short-running job injections. The impact ranges from around 3\% to 14\%, namely 1.8 - 8.4 minutes.

\begin{mybox}
\textbf{Observation 10.} Both small-sized and short-running capacity jobs can effectively utilize the idle computing resources in the capability system Theta, thereby improving its resource utilization. As more capacity jobs are injected into Theta, they can address temporal and spatial allocation gaps more effectively. Cori’s small-sized and short-running jobs can help Theta approach full utilization on multiple days.
\end{mybox}

\begin{mybox}
\textbf{Observation 11.}
Adding extra capacity jobs on Theta impacts the average wait time of Theta jobs,with the degree of impact depending on the types of injected jobs. The impact may be minor, for example, not more than 7\%, even when more than half a million capacity jobs are integrated into Theta. 
\end{mybox}

Figure~\ref{fig:injection_avg_wait_time_by_category} presents the impacts on various categories of Theta jobs concerning job size and runtime. 
When injecting small-sized jobs (W1-W2, adding approximately 1,500 extra jobs to Theta daily), we observe a notable impact on large-sized (1k+ job size) or long-running (runtime larger than 8.3 hours) Theta jobs, resulting in a wait time increase of over 10\%. 
However, certain Theta jobs, such as 128-node jobs or those with runtimes less than 4.17 hours, experience reduced wait times. For instance, Theta jobs lasting less than 2 hours experienced reduced wait times by up to 8.43\% when injecting Cori jobs consisting of 1-128 nodes with runtimes of less than 45 minutes (i.e., W1 and W2).
Our analysis suggests that this phenomenon occurs because backfilling Cori jobs may delay the arrival of large-sized Theta jobs, as depicted in Figure~\ref{fig:injection_avg_wait_time_by_category}(a). As a result, when a large-sized Theta job experiences a delay and reserves its resources, smaller Theta jobs (such as 128-node Theta jobs) can take advantage of this gap in scheduling, leading to a reduction in the average job wait time. 
We also observe that injecting a large number of small-sized Cori jobs (W3) can result in longer job wait times for all categories of Theta jobs. It's worth noting that W3 comprises over 800K jobs, translating to an average addition of over 2,200 Cori jobs into Theta per day.

\begin{mybox}
\textbf{Observation 12.}
Capability jobs on Theta, with varying sizes and runtimes, respond differently to the injection of capacity jobs. Small-sized jobs (1-128 nodes) from Cori, with runtimes not exceeding 45 minutes, generally have negligible or minimal effects on the wait times of most Theta jobs, except for those that are large-sized (exceeding 1,000 nodes) or long-running (over 8.3 hours). However, large-sized jobs or those with long runtimes experience a wait time increase of over 10\%.
\end{mybox}

\begin{mybox}
\textbf{Observation 13.} 
Injecting short-running capacity jobs, lasting less than 30 minutes, regardless of their size (ranging from 1 to 4,096 nodes), generally results in minimal or insignificant changes in job wait times across all categories of Theta jobs. 
Compared to injecting small-sized jobs, we find that injecting short-running capacity jobs with runtimes of less than 30 minutes is more advantageous in minimizing resource waste and reducing the impact on existing capability jobs within the Theta system.
\end{mybox}

In the case of injecting short-running jobs (W4-W6), we observe that the injection of W4 or W5 has no or only a minor impact on the job wait time of any Theta jobs, regardless of their sizes or runtimes.
Notably, these workload injections result in hundreds of capacity jobs of size 1-4,096 being added to Theta daily.
However, injecting W6 could lead to over a 10\% increase in wait time for Theta jobs, particularly for large-sized, long-running jobs.

Our analysis has addressed Q3 and Q4 listed in Section~\ref{method}. First, a capability system can leverage additional capacity jobs to fill its temporal and spatial allocation gaps, enhancing resource utilization. The injection of small-sized or short-running capacity jobs can contribute to the efficiency of the capability system. Second, strategically selecting capacity jobs to fill scheduling gaps is crucial for optimizing performance. In our case study, we find that injecting short-running jobs, rather than small-sized ones, is more beneficial. Short-running jobs of less than 30 minutes cause either no impact or only minor increases in the wait time of Theta jobs, irrespective of job size or runtime.

\section{Conclusion}\label{Conclusion}
Traditionally, capability-predominant and capacity-predominant workloads are scheduled and managed on separate computer clusters. These silos in the computing ecosystem have created barriers, leading to challenges such as resource underutilization, load imbalance, and cost inefficiency. Recognizing the limitations of siloed computing paradigms (capability vs capacity), we have explored the potential benefits and impacts that could be unlocked by creating a unified computing infrastructure. 

We first characterized two representative computing workloads: capability-predominant and capacity-predominant, from production systems.
Subsequently, we presented two integration analyses: (i) workload fusion, focusing on deploying a unified system for accommodating both capability-predominant and capacity-predominant workloads, and (ii) workload injection, leveraging capacity workloads to enhance resource utilization of the capability platform. 
Our trace-based integration analysis provided valuable insights into the potential scheduling and cost-effectiveness benefits of strategically integrating capability-predominant and capacity-predominant workloads. 

We will open-source the software and data utilized in this study. We hope our study encourages HPC researchers to promote unified computing in the community. 


\section*{Acknowledgment}
This work is supported in part by US National Science Foundation grants CCF-2413597, OAC-2402901 and Department of Energy under Contract No. DE-SC0024271.

\newpage
\bibliographystyle{ieeetr}
\bibliography{bib/IEEE}

\begin{thebibliography}{10}

\bibitem{theta}
ALCF, ``Theta.'' \url{https://www.alcf.anl.gov/theta}, 2024.

\bibitem{cori}
NERSC, ``{Cori}.''
  \url{"http://www.nersc.gov/users/computational-systems/cori/"}, 2024.

\bibitem{bicer2011framework}
T.~Bicer, D.~Chiu, and G.~Agrawal, ``A framework for data-intensive computing
  with cloud bursting,'' in {\em 2011 IEEE international conference on cluster
  computing}, (Austin, TX, USA), pp.~169--177, IEEE, 2011.

\bibitem{nair2010towards}
S.~K. Nair, S.~Porwal, T.~Dimitrakos, A.~J. Ferrer, J.~Tordsson, T.~Sharif,
  C.~Sheridan, M.~Rajarajan, and A.~U. Khan, ``Towards secure cloud bursting,
  brokerage and aggregation,'' in {\em 2010 eighth IEEE European conference on
  web services}, (Ayia Napa, Cyprus), pp.~189--196, IEEE, 2010.

\bibitem{fan2019scheduling}
Y.~Fan, Z.~Lan, P.~Rich, W.~E. Allcock, M.~E. Papka, B.~Austin, and D.~Paul,
  ``Scheduling beyond cpus for {HPC},'' in {\em Proceedings of the 28th
  International Symposium on High-Performance Parallel and Distributed
  Computing}, (Phoenix, Arizona, United States), p.~97–108, ACM, 2019.

\bibitem{mao2016resource}
H.~Mao, M.~Alizadeh, I.~Menache, and S.~Kandula, ``Resource management with
  deep reinforcement learning,'' in {\em Proceedings of the 15th ACM workshop
  on hot topics in networks}, (Atlanta, GA, USA), p.~50–56, ACM, 2016.

\bibitem{rlscheduler}
D.~Zhang, D.~Dai, Y.~He, F.~S. Bao, and B.~Xie, ``Rlscheduler: an automated hpc
  batch job scheduler using reinforcement learning,'' in {\em SC20:
  International Conference for High Performance Computing, Networking, Storage
  and Analysis}, (Virtual), pp.~1--15, ACM, 2020.

\bibitem{gunasekaran2019spock}
J.~R. Gunasekaran, P.~Thinakaran, M.~T. Kandemir, B.~Urgaonkar, G.~Kesidis, and
  C.~Das, ``Spock: Exploiting serverless functions for slo and cost aware
  resource procurement in public cloud,'' in {\em 2019 IEEE 12th international
  conference on cloud computing (CLOUD)}, (Milan, Italy), pp.~199--208, IEEE,
  2019.

\bibitem{shahrad2020serverless}
M.~Shahrad, R.~Fonseca, I.~Goiri, G.~Chaudhry, P.~Batum, J.~Cooke, E.~Laureano,
  C.~Tresness, M.~Russinovich, and R.~Bianchini, ``Serverless in the wild:
  Characterizing and optimizing the serverless workload at a large cloud
  provider,'' in {\em 2020 USENIX annual technical conference (USENIX ATC 20)},
  (Virtual), pp.~205--218, USENIX, 2020.

\bibitem{farahabady2013pareto}
M.~R.~H. Farahabady, Y.~C. Lee, and A.~Y. Zomaya, ``Pareto-optimal cloud
  bursting,'' {\em IEEE Transactions on Parallel and Distributed Systems},
  vol.~25, no.~10, pp.~2670--2682, 2013.

\bibitem{national2008potential}
N.~R. Council {\em et~al.}, {\em The potential impact of high-end capability
  computing on four illustrative fields of science and engineering}.
\newblock Virtual: National Academies Press, 2008.

\bibitem{mu2001utilization}
A.~W. Mu'alem and D.~G. Feitelson, ``Utilization, predictability, workloads,
  and user runtime estimates in scheduling the ibm sp2 with backfilling,'' {\em
  IEEE Transactions on Parallel and Distributed Systems}, vol.~12, no.~6,
  pp.~529--543, 2001.

\bibitem{allcock2017experience}
W.~Allcock, P.~Rich, Y.~Fan, and Z.~Lan, ``Experience and practice of batch
  scheduling on leadership supercomputers at argonne,'' {\em Job Scheduling
  Strategies for Parallel Processing}, vol.~10773, pp.~1--24, 2017.

\bibitem{emeras2014analysis}
J.~Emeras, C.~Ruiz, J.-M. Vincent, and O.~Richard, ``Analysis of the jobs
  resource utilization on a production system,'' in {\em JSSPP 2013}, (Boston,
  MA, USA), pp.~1--21, Springer, 2014.

\bibitem{you2013comprehensive}
H.~You and H.~Zhang, ``Comprehensive workload analysis and modeling of a
  petascale supercomputer,'' in {\em Job Scheduling Strategies for Parallel
  Processing: 16th International Workshop, JSSPP 2012, Shanghai, China, May 25,
  2012. Revised Selected Papers 16}, (Shanghai, China), pp.~253--271, Springer,
  2013.

\bibitem{rodrigo2018towards}
G.~P. Rodrigo, P.-O. {\"O}stberg, E.~Elmroth, K.~Antypas, R.~Gerber, and
  L.~Ramakrishnan, ``Towards understanding hpc users and systems: a nersc case
  study,'' {\em Journal of Parallel and Distributed Computing}, vol.~111,
  pp.~206--221, 2018.

\bibitem{cortez2017resource}
E.~Cortez, A.~Bonde, A.~Muzio, M.~Russinovich, M.~Fontoura, and R.~Bianchini,
  ``Resource central: Understanding and predicting workloads for improved
  resource management in large cloud platforms,'' in {\em Proceedings of the
  26th Symposium on Operating Systems Principles}, (Shanghai, China),
  pp.~153--167, ACM, 2017.

\bibitem{li2023analyzing}
J.~Li, G.~Michelogiannakis, B.~Cook, D.~Cooray, and Y.~Chen, ``Analyzing
  resource utilization in an hpc system: A case study of nersc’s
  perlmutter,'' in {\em International Conference on High Performance
  Computing}, (Bangalore, India), pp.~297--316, Springer, 2023.

\bibitem{patel2020job}
T.~Patel, Z.~Liu, R.~Kettimuthu, P.~Rich, W.~Allcock, and D.~Tiwari, ``Job
  characteristics on large-scale systems: long-term analysis, quantification,
  and implications,'' in {\em SC20: International conference for high
  performance computing, networking, storage and analysis}, (Virtual),
  pp.~1--17, ACM, 2020.

\bibitem{ferreira2021workflows}
R.~Ferreira~da Silva, K.~Chard, H.~Casanova, D.~Laney, D.~H. Ahn, S.~Jha, W.~E.
  Allcock, G.~Bauer, D.~Duplyakin, B.~Enders, {\em et~al.}, ``Workflows
  community summit: Tightening the integration between computing facilities and
  scientific workflows,'' tech. rep., Oak Ridge National Lab.(ORNL), Oak Ridge,
  TN (United States), 2021.

\bibitem{li2022ai}
B.~Li, R.~Arora, S.~Samsi, T.~Patel, W.~Arcand, D.~Bestor, C.~Byun, R.~B. Roy,
  B.~Bergeron, J.~Holodnak, {\em et~al.}, ``Ai-enabling workloads on
  large-scale gpu-accelerated system: characterization, opportunities, and
  implications,'' in {\em 2022 IEEE International Symposium on High-Performance
  Computer Architecture (HPCA)}, (Seoul, South Korea), pp.~1224--1237, IEEE,
  2022.

\bibitem{plan_based}
X.~Zheng, Z.~Zhou, X.~Yang, Z.~Lan, and J.~Wang, ``Exploring plan-based
  scheduling for large-scale computing systems,'' in {\em 2016 IEEE
  International Conference on Cluster Computing (CLUSTER)}, (Taipei, Taiwan),
  pp.~259--268, IEEE, 2016.

\bibitem{fan2021deep}
Y.~Fan, Z.~Lan, T.~Childers, P.~Rich, W.~Allcock, and M.~E. Papka, ``Deep
  reinforcement agent for scheduling in hpc,'' in {\em 2021 IEEE International
  Parallel and Distributed Processing Symposium (IPDPS)}, (Virtual),
  pp.~807--816, IEEE, 2021.

\bibitem{przybylski2022using}
B.~Przybylski, M.~Pawlik, P.~{\.Z}uk, B.~Lagosz, M.~Malawski, and K.~Rzadca,
  ``Using unused: non-invasive dynamic faas infrastructure with hpc-whisk,'' in
  {\em SC22: International Conference for High Performance Computing,
  Networking, Storage and Analysis}, (Dallas, TX, USA), pp.~1--15, ACM, 2022.

\bibitem{roy2022mashup}
R.~B. Roy, T.~Patel, V.~Gadepally, and D.~Tiwari, ``Mashup: making serverless
  computing useful for hpc workflows via hybrid execution,'' in {\em
  Proceedings of the 27th ACM SIGPLAN Symposium on Principles and Practice of
  Parallel Programming}, (Seoul, South Korea), pp.~46--60, ACM, 2022.

\bibitem{du2019feasibility}
R.~Du, J.~Shi, J.~Zou, X.~Jiang, Z.~Sun, and G.~Chen, ``A feasibility study on
  workload integration between ht-condor and slurm clusters,'' {\em EPJ Web of
  Conferences}, vol.~214, p.~08004, 2019.

\bibitem{lahiff2020running}
A.~Lahiff, S.~de~Witt, M.~Caballer, G.~La~Rocca, S.~Pamela, and D.~Coster,
  ``Running htc and hpc applications opportunistically across private, academic
  and public clouds,'' {\em EPJ Web of Conferences}, vol.~245, p.~07032, 2020.

\bibitem{guo2014cost}
T.~Guo, U.~Sharma, P.~Shenoy, T.~Wood, and S.~Sahu, ``Cost-aware cloud bursting
  for enterprise applications,'' {\em ACM Transactions on Internet Technology
  (TOIT)}, vol.~13, no.~3, pp.~1--24, 2014.

\bibitem{guo2012seagull}
T.~Guo, U.~Sharma, T.~Wood, S.~Sahu, and P.~Shenoy, ``Seagull: Intelligent
  cloud bursting for enterprise applications,'' in {\em 2012 USENIX Annual
  Technical Conference (USENIX ATC 12)}, (Boston, MA, USA), pp.~361--366,
  USENIX, 2012.

\bibitem{KNL}
A.~Sodani, ``Knights landing (knl): 2nd generation intel{\textregistered} xeon
  phi processor,'' in {\em 2015 IEEE Hot Chips 27 Symposium (HCS)}, (Cupertino,
  CA, USA), pp.~1--24, IEEE, 2015.

\bibitem{Haswell}
P.~Hammarlund, A.~J. Martinez, A.~Bajwa, D.~L. Hill, E.~Hallnor, H.~Jiang,
  M.~Dixon, M.~Derr, M.~Hunsaker, R.~Kumar, {\em et~al.}, ``4th generation
  intel core processor, codenamed haswell,'' 2013.

\bibitem{CQSim}
SPEAR-UIC, ``{CQSim}.'' \url{"https://github.com/SPEAR-UIC/CQSim"}, 2024.

\bibitem{SWF}
D.~Talby, ``{Parallel Workload Archive}.''
  \url{"https://www.cs.huji.ac.il/labs/parallel/workload/swf.html"}, 2024.

\bibitem{WFP_tang_2011}
W.~Tang, Z.~Lan, N.~Desai, D.~Buettner, and Y.~Yu, ``Reducing fragmentation on
  torus-connected supercomputers,'' in {\em 2011 IEEE International Parallel \&
  Distributed Processing Symposium}, (Anchorage, Alaska, USA), pp.~828--839,
  IEEE, 2011.

\bibitem{yang-sc13}
X.~Yang, Z.~Zhou, S.~Wallace, Z.~Lan, W.~Tang, S.~Coghlan, and M.~E. Papka,
  ``{Integrating dynamic pricing of electricity into energy aware scheduling
  for HPC systems},'' in {\em SC '13: Proceedings of the International
  Conference on High Performance Computing, Networking, Storage and Analysis},
  (Denver, Colorado, USA), pp.~1--11, ACM, 2013.

\bibitem{rountree2016}
N.~Gholkar, F.~Mueller, and B.~Rountree, ``A power-aware cost model for hpc
  procurement,'' in {\em 2016 IEEE International Parallel and Distributed
  Processing Symposium Workshops (IPDPSW)}, 2016.

\end{thebibliography}


\end{document}